\documentstyle[12pt]{article}   

\large
\newcommand{\beq}{\begin{equation}}
\newcommand{\eeq}{\end{equation}}

\makeatletter
\@addtoreset{equation}{section}
\makeatother

\newtheorem{Theorem}{Theorem}[section]
\newtheorem{Definition}{Definition}[section]
\newtheorem{Lemma}{Lemma}[section]
\newtheorem{Corollary}{Corollary}[section]

\def\be{\begin{equation}}
\def\ee{\end{equation}}
\def\ba{\begin{eqnarray}}
\def\ea{\end{eqnarray}}

\def\agb{{\overline {{\cal A}/{\cal G}}}}

\def\a{\alpha}

\def\Comp{{\mathchoice
{\setbox0=\hbox{$\displaystyle\rm C$}\hbox{\hbox to0pt
{\kern0.4\wd0\vrule height0.9\ht0\hss}\box0}}
{\setbox0=\hbox{$\textstyle\rm C$}\hbox{\hbox to0pt
{\kern0.4\wd0\vrule height0.9\ht0\hss}\box0}}
{\setbox0=\hbox{$\scriptstyle\rm C$}\hbox{\hbox to0pt
{\kern0.4\wd0\vrule height0.9\ht0\hss}\box0}}
{\setbox0=\hbox{$\scriptscriptstyle\rm C$}\hbox{\hbox to0pt
{\kern0.4\wd0\vrule height0.9\ht0\hss}\box0}}}}
\def\Co{{\mathchoice
{\setbox0=\hbox{$\displaystyle\rm C$}\hbox{\hbox to0pt
{\kern0.4\wd0\vrule height0.9\ht0\hss}\box0}}
{\setbox0=\hbox{$\textstyle\rm C$}\hbox{\hbox to0pt
{\kern0.4\wd0\vrule height0.9\ht0\hss}\box0}}
{\setbox0=\hbox{$\scriptstyle\rm C$}\hbox{\hbox to0pt
{\kern0.4\wd0\vrule height0.9\ht0\hss}\box0}}
{\setbox0=\hbox{$\scriptscriptstyle\rm C$}\hbox{\hbox to0pt
{\kern0.4\wd0\vrule height0.9\ht0\hss}\box0}}}}
\def\Rl{{\mathchoice
{\setbox0=\hbox{$\displaystyle\rm R$}\hbox{\hbox to0pt
{\kern0.4\wd0\vrule height0.9\ht0\hss}\box0}}
{\setbox0=\hbox{$\textstyle\rm R$}\hbox{\hbox to0pt
{\kern0.4\wd0\vrule height0.9\ht0\hss}\box0}}
{\setbox0=\hbox{$\scriptstyle\rm R$}\hbox{\hbox to0pt
{\kern0.4\wd0\vrule height0.9\ht0\hss}\box0}}
{\setbox0=\hbox{$\scriptscriptstyle\rm R$}\hbox{\hbox to0pt
{\kern0.4\wd0\vrule height0.9\ht0\hss}\box0}}}}

\title{Quantum Spin Dynamics (QSD) II}
\author{T. Thiemann\thanks{thiemann@math.harvard.edu} \\
       Physics Department, Harvard University, \\
       Cambridge, MA 02138, USA}
\date{{\small Preprint HUTMP-96/B-352}}

\begin{document}

\maketitle

\begin{abstract}
We continue here the analysis of the previous paper of the Wheeler-DeWitt 
constraint 
operator for four-dimensional, Lorentzian, non-perturbative, canonical 
vacuum quantum gravity in the continuum.
In this paper we derive the complete kernel, as well as a physical inner 
product on it, for a non-symmetric version of the Wheeler-DeWitt operator. 
We then define a symmetric version of the Wheeler-DeWitt operator.
For the Euclidean Wheeler-DeWitt operator as well as for the generator
of the Wick transform from the Euclidean to the Lorentzian regime we prove
existence of self-adjoint extensions and based on these we present a method
of proof of self-adjoint extensions for the Lorentzian operator. 
Finally we comment on the status of the Wick rotation transform in the light
of the present results.
\end{abstract}

\section{Complete physical Hilbert space and observables}

In this section we will compute the complete kernel of both the 
Diffeomorphism and the non-symmetric Euclidean and Lorentzian 
Hamiltonian constraint (for the symmetric Hamiltonian operator, see the 
next section). 
The kernel turns out to be spanned by distributions which do not only 
involve cylindrical 
functions which live on at most two-valent graphs or on graphs 
containing vertices with arbitrary valence but such that at each vertex 
the tangents of incident edges are co-planar. These solutions involve 
vertices of arbitrary valence and intersection characteristics, do take the 
curvature term $F_{ab}$ of the classical constraint fully into account 
and are not necessarily annihilated by the volume operator. Also they 
are sensitive to whether they belong to the kernel of the Euclidean 
or Lorentzian Hamiltonian constraint.\\
This space of distributional solutions inherits a
natural inner product coming from $\cal H$ via the group averaging method
and it turns out that it coincides with the one given in \cite{18}. 
Furthermore, we will define the notion of an observable and give 
explicit, non-trivial examples of those.\\
The key observation is the following :\\
Consider the action of $\hat{H}^E(N)$ on a spin-network state 
$T_{\gamma,\vec{j},\vec{c}}$ defined on a graph $\gamma$. Then 
$\hat{H}^E(N)T_{\gamma,\vec{j},\vec{c}}$ can be written as a finite linear
combination of spin-network states defined on graphs $\gamma_I$ where
$\gamma\subset\gamma_I$ and $a_I:=\gamma_I-\gamma$ is precisely one of the 
arcs $a_{ij}(\Delta)$. Moreover, $a_I$ carries spin $j_I=1/2$ because the 
arcs $a_{ij}(\Delta)$ do not appear in $\gamma$ but they appear in 
$\hat{H}^E(N)$ through the fundamental representation of $SU(2)$. The arcs
$a_I$ are special edges of $\gamma_I$ in the following sense.
\begin{Definition} \label{def3}
1) A vertex $v$ of a graph $\gamma$ is called extraordinary provided that\\
i) it is tri-valent,\\
ii) it is the intersection of precisely two analytic curves 
$c,c'\subset\gamma$, that is,
$v=c\cap c'$, such that $v$ is an endpoint of $c$ but not of $c'$.\\
2) An edge $e$ of a graph $\gamma$ is called extraordinary provided that\\
i) its endpoints $v_1,v_2$ are both extraordinary vertices of $\gamma$,\\
ii) there is an at least trivalent vertex $v$ of $\gamma$ which is such that
at least three edges incident at it have linearly independent tangents at
$v$ and there are two edges $s_1,s_2\subset\gamma$ respectively which connect
$v$ and $v_1,v_2$ respectively and which have linearly independent 
tangents at $v$. We will call $v$ the typical vertex associated with 
$e$. 
\end{Definition}
In other words, if $e_1,e_2$ is the connected part of the intersection of 
the analytic extensions of $s_1,s_2$ with $\gamma$ that contains $s_1,s_2$
then $e_1\cup e_2\cup e$ looks like the graph picturized as $\forall$. It is 
easy to check that
$a_I$ is an extraordinary edge for $\gamma_I$ and so a rough description 
of the action of $\hat{H}^E(N)$ is by saying that it admits a 
decomposition into spin-network states defined on graphs which differ by one 
extraordinary edge with spin $1/2$ compared to the original graph.\\
Next let us look at $\hat{K}$. Since $\hat{K}\propto[\hat{V},\hat{H}^E(1)]$
it follows that $\hat{K}$ has the same property. Finally, since 
$s_i(\Delta)$ are not extraordinary edges of a given graph $\gamma$ it 
follows that the action of $\hat{T}(N)$ can be described by saying 
that it admits a decomposition into spin-network states defined 
on graphs which differ by two, necessarily 
disjoint, extraordinary edges with spin $1/2$ compared to the original graph.
This is because $\hat{T}(N)$ contains two factors of $\hat{K}$.\\
\begin{Definition} \label{def4}
i) A spin-net is a pair $w=(\gamma,\vec{j})$ consisting of a graph $\gamma
\in\Gamma$ and a colouring of the edges of $\gamma$ with spins $j>0$
such that the set of vertex contractors compatible with the data 
$\gamma,\vec{j}$ is not empty. We will denote the set of all spin-nets 
by W.\\
ii) The subset $W_0\subset W$ is defined to be the set of all 
$(\gamma,\vec{j})\in W$ where $\gamma$ is a 
piecewise analytic graph all of whose extraordinary edges carry a spin 
$j>1/2$. We also set $\overline{W}_0:=W-W_0$.\\
iii) Given $w=(\gamma,\vec{j})\in W$ there exists a unique spin-net 
$w_0(w)=((\gamma_0(\gamma),\vec{j}_0(\vec{j}))$, called the source of $w$
and which is defined by the subsequent algorithm :\\
First, let $\tilde{\gamma}$ be a copy of $\gamma$ which we dye in 
white.\\
If $w\not\in W_0$ remove all the extraordinary edges $e$ of $\gamma$ which 
carry spin $1/2$ in $\gamma$ to obtain a 
graph $\gamma'$. Now, if $s_1,s_2$ are the segments of $\gamma$ which 
connect the extraordinary edge $e$ with its typical vertex then 
dye $s_1,s_2$ black in $\tilde{\gamma}$ (no matter which dye they had 
before) to produce $\tilde{\gamma}'$. 
Iterate the procedure with $\gamma',\tilde{\gamma}'$ instead of 
$\gamma,\tilde{\gamma}$.
The procedure must come to an end after a finite number of steps because 
$\gamma$ had only a finite number of edges. The final $\gamma'$ is the 
searched for $\gamma_0(\gamma)$ which by construction is unique. Its
colouring $\vec{j}_0$ is obtained as follows : 
Each edge $e$ of $\gamma_0(\gamma)$ has a 
finite segment $s$ which is dyed in white in the final $\tilde{\gamma}$ and 
which 
belongs to an edge $e'$ of $\gamma$. We define $\vec{j}_0(\vec{j})$ by
requiring that the colour of $e$ is the same as that of $e'$.   
It is clear that the pair $(\gamma_0,\vec{j}_0)$ defines an element of
$W_0$ : it is is an element of $W$ because the space of vertex contractors 
associated with a trivalent vertex as that given by the endpoints of an 
extraordinary edge is one-dimensional and that it lies in $W_0$ follows 
from the construction.
\end{Definition}
In order to characterize the complete set of solutions we need one more 
definition.
\begin{Definition} \label{def5}
a) Let $w_0=(\gamma_0,\vec{j}_0)\in W_0$. 
We define inductively finite sets of spin-nets $w=(\gamma,\vec{j})\in
\overline{W}_0$ with source $w_0$ as follows :\\
1) Let $W^{(0)}(w_0):=\{w_0\}$.\\
2) Given $W^{(n)}(w_0)$ take any $(\gamma,\vec{j})\in
W^{(n)}(w_0)$ and construct elements $(\gamma',\vec{j}')$ of 
$W^{(n+1)}(w_0)$ as follows : add precisely one 
more extraordinary edge $e$ to $\gamma$ in all possible, topologically
inequivalent, ways. Furthermore, if
$v$ is the typical vertex for $e$ and if $e_i=s_i\circ s_i',i=1,2,
s_i,s_i'\not=\emptyset$ carries spin $j_i$ where $s_1,s_2$ connect $v$ 
to the endpoints of $e$ then we define up to four colourings for 
$\gamma\cup e$ as follows :\\
i) The extraordinary edge $e$ is coloured with spin $1/2$.\\
ii) $s_i'$ is coloured with spin $j_i$ as before.\\
iii) $s_i$ is coloured with spin $j_i':=j_i\pm 1/2$. \\
iv) The edges of $\gamma-\{e_1,e_2\}$ carry the same spin as in $\gamma$.\\
v) from the colourings of $\gamma\cup e$ so obtained we keep only those 
which admit a non-empty set of vertex contractors.\\
vi) Define $\gamma':=\gamma\cup e,(\gamma-s_1)\cup e,(\gamma-s_2)\cup e,
(\gamma-s_1-s_2)\cup e$ if $(j_1',j_2')$ is $(\not=0,\not=0),
(0,\not=0),(\not=0,0),(0,0)$ respectively.\\
The set of data $(\gamma',\vec{j}')$ (at most four) for each
$(\gamma,\vec{j})$ and for each $e$ extraordinary for $\gamma$ so obtained 
comprises the set $W^{(n+1)}(w_0)$.\\
The finite set $W^{(n)}(w_0)$ will be called the set of 
derived spin-nets of level $n$ with source  $w_0$.\\
b) We will denote the associated set of equivalence classes of spin-nets 
under diffeomorphisms by $[W^{(n)}(w_0)]$ which itself, of course, depends
only on the equivalence class $[w_0]$ of $w_0$.
\end{Definition}
Notice that no graph involved in the derived spin-nets can get get 
disconnected because there must have been $n\ge 3$ edges involved at the 
typical vertex under question. Therefore the combination $j_1'=j_2'=0$ can
actually only occur for $n\ge 4$ because of condition $a),v)$. It follows 
that we produce only vertices with minimal valence two but then at the 
next level this is not a typical vertex any longer.\\
It is therefore clear that for each $w\in\overline{W}_0$ there is 
precisely one $n>0$ such that $w\in W^{(n)}(w_0(w))$. In other words,
$\overline{W}_0$ can be derived from $W_0$.\\
Finally, we recall the definition of diffeomorphism invariant state
\cite{18}.
\begin{Definition} \label{def6}
i) Let $T_{\gamma,\vec{j},\vec{c}}$ be a spin-network state. Its group 
average is defined by the following well-defined distribution on
$\Phi$
\be \label{41}
T_{[\gamma],\vec{j},\vec{c}}:=\sum_{\gamma'\in[\gamma]}
T_{\gamma',\vec{j},\vec{c}}
\ee
where $[\gamma]$ denotes the orbit of $\gamma$ under smooth 
diffeomorphisms of $\Sigma$ which preserve analyticity of $\gamma$.\\
ii) The group average $[f]$ of any cylindrical function $f$ is defined by 
first decomposing it into spin-network states and then averaging each of 
the spin-network states separately.
\end{Definition}
As shown in \cite{18}, the distributions of the form $\Psi:=[f]$ provide 
the general solution to the diffeomorphism constraint. Moreover one can 
show that 
\be \label{42}
T_{[\gamma],\vec{j},\vec{c}}(T_{\gamma',\vec{j}',\vec{c}'})
=\chi_{[\gamma]}(\gamma')\delta_{\vec{j},\vec{j}'}\delta_{\vec{c},\vec{c'}}
\ee
where $\chi$ denotes the characteristic function.
We are now ready to define the complete set of simultaneous solutions to the 
Diffeomorphism constraint and to the non-symmetric Lorentzian Hamiltonian 
constraint as well as a physical inner product thereon. 
\begin{Theorem} \label{th10}
Each distributional solution to all constraints of Lorentzian quantum gravity
is a finite linear combination of states $\Psi$ of the following two 
types :\\ 
Type I)\\
$\Psi=[f]$ where $f$ is an arbitrary linear combination of spin-network 
states based on spin-nets $w_0\in W_0$.\\
Type II)\\
$\Psi=[f]$ where $f$ is a certain linear combination of spin-network 
states which are constructed from spin-nets in $\overline{W}_0$. We
will characterize this linear combination precisely in the course of the
proof.
\end{Theorem}
Proof :\\
Clearly both types of vectors solve the diffeomorphism constraint.\\
The basic observation is that if we have a spin-network state 
$T_{\gamma,\vec{j},\vec{c}}$ then $\hat{H}^E(N)T_{\gamma,\vec{j},\vec{c}}$
is a linear combination of spin-network states 
$T_{\gamma',\vec{j}',\vec{c}'}$ where $\gamma'$ has precisely one edge 
$e$ more than $\gamma$, moreover, 
$e$ is extraordinary edge coloured with spin $1/2$. Likewise, 
$\hat{T}(N)T_{\gamma,\vec{j},\vec{c}}$ is a linear combination of such
spin-network states where $\gamma'$ has precisely two disjoint edges $e,f$ 
more than
$\gamma$, where at least one of them, say $e$, is an extraordinary edge for 
$\gamma'$ coloured with spin $1/2$ and where at least one of them, say $f$,
is an extraordinary edge for $\gamma'-e$ coloured with spin $1/2$.
It follows that necessarily $\hat{H}(N)T_{\gamma,\vec{j},\vec{c}}$ is
a linear combination of spin-network states which are compatible with 
spin-nets $w\in\overline{W}_0$.

By definition of a solution of the Hamiltonian constraint we have to check
that $\Psi(\hat{H}(N)f)=0$ for all lapses $N$ and all cylindrical $f$
which is clearly equivalent to showing that $\Psi(\hat{H}(N)
T_{\gamma,\vec{j},\vec{c}})=0$ for all $N$ and 
all $T_{\gamma,\vec{j},\vec{c}}$.

Now let first $\Psi=[f]$ be of type I. The condition is trivially met because
even if $f$ contains a spin-network state $T_{\gamma\ast,\vec{j}^\ast,
\vec{c}^\ast}$ based on a graph $\gamma^\ast$ which is 
diffeomorphic to a graph $\gamma'$ where $T_{\gamma',\vec{j}',\vec{c}'}$
is one of the spin-network states into which 
$\hat{H}(N)T_{\gamma,\vec{j},\vec{c}}$ can be decomposed, the spin vectors
$\vec{j}^\ast,\vec{j}'$ are necessarily different in at least one 
extraordinary
edge which carries spin $1/2$ in $\gamma'$ but not in $\gamma$ and so the 
inner product
(\ref{42}) vanishes. The solutions of type I are in a sense trivial 
because every operator which extends the graph of a function cylindrical 
with respect to it by edges of particular topology and spin value will 
have the same type of solutions.

Consider now solutions of type II).
Let $f=\sum_T a^{(n)}_{[T]}([w_0]) T$ where the sum
extends over 1) all spin-nets $w\in W^{(n)}(w_0)$ for some 
$w_0=(\gamma_0,\vec{j}_0)\in W_0$ and some $n>0$ and 2) over all 
spin-network states $T$ compatible with precisely one of these w (we will 
call this set $S^{(n)}(w_0))$. Now, by the explicit
expression of $\hat{H}(N)$ \cite{0},(5.3), it follows that $\hat{H}^E(N)$
maps precisely all $T\in S^{(n-1)}(w_0)$
into linear combinations of spin-network states which are 
diffeomorphic with some of the elements  $T'\in W^{(n)}(w_0)$ and no other 
spin-network states do have this property. Likewise, $\hat{T}(N)$
maps precisely all $T\in S^{(n-2)}(w_0)$
into linear combinations of spin-network states which are 
diffeomorphic with elements  $T'\in W^{(n)}(w_0)$ and no other 
spin-network states do have this property. It follows that we have
matrices $m^{(n)}_{[T],[T']}([w_0],[v])$ such that 
\ba
\hat{H}^E(N)T&=&\sum_{v\in V(\gamma_0),T'\in S^{(n)}([w_0])}
N_v m^{(n)}_{[T],[T']}([w_0],[v]) T'\mbox{ for }T\in S^{(n-1)}(w_0)
\nonumber\\
\hat{T}(N)T&=&\sum_{v\in V(\gamma_0),T'\in S^{(n)}(w_0)}
N_v m^{(n)}_{[T],[T']}([w_0],[v]) T'\mbox{ for }T\in S^{(n-2)}(w_0)\;.
\nonumber
\ea
Here we mean by $T'$ one of the representants of the diffeomorphism class 
of vectors into which $T$ is mapped.
Notice that the matrices $m$ are diffeomorphism invariant which follows
from the fact that they can only depend on the $\vec{j},\vec{c}$ involved.
It follows that $[f]$ is a solution if and only if
\be \label{43}
\sum_{T'\in S^{(n)}(w_0)} 
\bar{a}^{(n)}_{[T']}([w_0]) m^{(n)}_{[T],[T']}([w_0],[v])=0\;\forall
\;T\in S^{(n-1)}(w_0)\cup S^{(n-2)}(w_0),\;v\in V(\gamma_0).
\ee
This is the condition that we looked for.\\
Since the members of all the $S^{(n)}(w_0)$ for all $w_0$ obviously comprise 
all the spin-network states compatible with any $w\in\overline{W}_0$ 
it follows that we have found the general solution.
$\Box$\\
\begin{Corollary}
Every solution of the Lorentzian Hamiltonian constraint solves the
Euclidean Hamiltonian constraint as well.
\end{Corollary}
This follows obviously from the proof given above because the two
parts $\hat{H}^E,\hat{T}$ of $\hat{H}$ need to vanish separately.
It follows that Lorentzian solutions are rather special elements of the 
bigger set of Euclidean solutions.\\
\\
A few remarks are in order :
\begin{itemize}
\item[0)] Notice that the Diffeomorphism constraint moves the graph of a 
spin-network state but leaves the spin data $\vec{j},\vec{c}$ invariant.
On the other hand, the Hamiltonian constraint is only a condition on the
spin-data. It is here where the dynamics is encoded. It is interesting that
the two constraints effectively act on different, nicely split, labels of
a spin-network state. The solutions of type II) are neatly labelled by
the $[W^{(n)}(w_0)]$, that is by the diffeomorphism classes $[W_0]$ and 
by the number $n$, which can roughly be interpreted as the number of times
that $\hat{H}^E(N)$ acts on an element $w_0$ of $W_0$.
\item[1)] Notice that if we wished to solve the Hamiltonian constraint {\em 
before} the diffeomorphism constraint then we could actually do so : 
Theorem \ref{th10} would still hold, we just need to drop the group 
averaging. Remarkably, the solutions $\psi$ are then not even
distributional, they are elements of $\Phi$.
\item[2)] Let us then assume that we solve the Hamiltonian constraint 
before the diffeomorphism constraint. How do our solutions then compare 
with those found in the literature \cite{26,27} ? The authors of those 
papers try to compute the kernel of $\hat{H}^E(N)$, that is, the space of 
solutions to the ordinary eigenvector equation $\hat{H}^E(N)\psi=0$, albeit
only for the Euclidean constraint. That is, the point $\lambda=0$ of the 
spectrum is analyzed by treating it as a part of the {\em point} spectrum 
(that is, there exists an eigenvector, which, in particular, is 
square-integrable, with eigenvalue $0$).\\
Now, although we do not have a complete proof, the fact that $\hat{H}^E(N)$ 
enlarges the graph
of a cylindrical function that it acts on seems to exclude the possibility
of a large enough kernel of $\hat{H}^E(N)$ when $0$ is considered as 
a part of the point spectrum. In a sense it is 
very similar to trying to find the eigenvectors of the creation operator 
$\hat{a}^\dagger$ of the harmonic oscillator Hamiltonian. The only 
solution ($0$) is trivial. The only zero eigenvectors which we find in 
our approach seem to be 
related to the solutions found in \cite{26,27} : they are spanned by 
functions cylindrical with respect to any graph of arbitrarily high valence 
but such that the tangents of all edges incident at any of its vertices 
are co-planar. We conjecture that this is the complete kernel 
corresponding to the eigenvalue zero.
It is clearly too small because these vectors are already 
annihilated by the volume operator, i.e. they do 
not take the curvature $F_{ab}$ (except for its anti-symmetry in $a,b$) into
account and so are not specific for $\hat{H}^E(N),\hat{H}(N)$. On the other 
hand,
they are the first known non-distributional {\em rigorous} solutions also 
for the {\em Lorentzian}  
Hamiltonian constraint in the continuum (the Lorentzian constraint defined 
on the lattice considered in \cite{28} blows up on those states
because this operator is only defined on states with finite volume). This 
is because both of $\hat{V},\hat{H}^E$ and therefore also $\hat{K}$
annihilate such vectors.\\
Therefore one is naturally led to the viewpoint that $0$ should not be 
considered as a part of the point spectrum : The point $0$ of the spectrum
is singled out in the sense that even though there maybe zero eigenvectors,
they are clearly not in the range of $\hat{H}(N)$ (which is not the 
case for eigenvalues different from zero). So, neglecting the fact that
$0$ {\em is} an eigenvalue we may treat $0$ as part of the residual 
spectrum, that is, the range of $\hat{H}(N)$ is 
not dense in $\cal H$ (notice that by the usual definition the point and 
residual spectra are automatically disjoint). The kernel of $\hat{H}(N)$
should then be considered as a subspace of the Hilbert space dual of 
$\cal H$, that is we look for $\psi\in{\cal H}'=\cal H$ such that
$\psi(\hat{H}(N) f)=0$ for all $\psi\in \cal H$ and so we 
automatically capture the zero eigenvectors as solutions $\psi$. The last
condition is equivalent to $\hat{H}(N)^\dagger \psi=0$, in other words,
treating $0$ as part of the residual spectrum of $\hat{H}(N)$ is equivalent
to treating it as part of the point spectrum of $\hat{H}(N)^\dagger$ in 
order to get the same kernel
(recall that in general, at least for bounded operators $\hat{O}$, the 
residual spectrum of $\hat{O}$ and the point spectrum of $\hat{O}^\dagger$ 
coincide).
Notice that it was precisely the fact that the kernel of $\hat{H}(N)$
is not dense in $\cal H$  which was exploited in Theorem \ref{th10} :
since $\hat{H}(N)$ extends the graph of a spin-network state to one with 
vertices and edges of 
a special kind and colours its edges in a particular way, its range is not 
dense. Speaking again in terms of 
an analogy with the harmonic oscillator, the adjoint of the creation 
operator, the annihilation operator, has a rich point spectrum, the
corresponding eigenvectors are even overcomplete.
\end{itemize}
\begin{Definition}
i) Consider the vector space of solutions ${\cal V}\subset\Phi'$ and 
complete it with 
respect to the inner product defined (and extended by sesquilinearity) by
\[ <[f],[f']>_{phys}:=[f](f')\;. \]
where $f,f'$ are any to cylindrical functions.
The resulting Hilbert space is called the physical Hilbert space
${\cal H}_{phys}$.\\
ii) An observable $[\hat{O}]$ is defined to be a self-adjoint operator on 
${\cal H}_{phys}$, densely defined on $\cal V$. Alternatively,
it is a self-adjoint operator densely defined on $\Phi$ such that its
extension to $\Phi'$ leaves $\cal V$ invariant.
\end{Definition}
A trivial example of an observable is the projector to the type
I solutions. That is, viewn as an operator on $\cal H$
we define for any function $f$ cylindrical with respect to a spin-net
$w=(\gamma,\vec{j})$
that $\hat{O}f=0$ if $w\not\in W_0$ and $\hat{O}f=f$ otherwise.
$\hat{O}$ is therefore densely defined and it is easy to see that it
is self-adjoint. It preserves solutions because if $\psi(\hat{H}(N)f)=0$
for all $f$ then clearly $\psi(\hat{H}(N)\hat{O}f)=0$ and since 
$\hat{O}\hat{H}(N)f=0$, trivially $\psi(\hat{H}(N)\hat{O}f)=0$.
An integral kernel representation of $\hat{O}$ is given by
$O(A,B)=\sum_T T(A)\bar{T}(B)$ where the sum is over all spin-network
states compatible with respect to webs in $W_0$. Viewn as an operator 
defined on
$\cal V$ we merely need to rearrange the last sum and collect sums over 
diffeomorphic graphs into the group average.

\section{Method to compute $a^{(n)}_{[T']}([w_0])$}

The precise computation of the coefficients $a^{(n)}_{[T]}(w_0)$ is 
straightforward but rather tedious. We will lay here the computational
foundations of an efficient computer code to obtain them. The 
details of the method 
are identical to those displayed in \cite{ThieVol,14} and will not be 
repeated here.\\
We consider the matrix elements of the volume operator on {\em 
extended} spin-network functions as known through (\cite{ThieVol}). By
an extended spin-network function we mean a function of the form
$T_{\gamma,\vec{j},\vec{c}}$ as before, the difference being that 
each $c_v$ of $\vec{c}=(c_v)_{v\in V(\gamma)}$ now maybe a projector on
a non-trivial irreducible representation of $SU(2)$, that is, the state 
is not gauge invariant.\\
Let $T_{\gamma,\vec{j},\vec{c}}$ be an extended spin-network function.
The operators $\hat{H}^E(N),\hat{T}(N)$ are gauge invariant but in applying
the volume operator involved in them we need extended spin-networks.
Consider first $\hat{H}^E_v$ which contains terms of the form
$\mbox{tr}([h_\alpha-h_\alpha^{-1}]h_s^{-1}\hat{V}h_s)$ where $s$ is a 
segment of an edge $e$ of $\gamma$ starting at $v$ and $\alpha$ is a loop 
based
at $v$ of the form $s'\circ a\circ(s^{\prime\prime})^{-1}$ where also
$s',s^{\prime\prime}$ are segments of edges $e',e^{\prime\prime}$ of 
$\gamma$ incident at $v$.
In order to compute the action of $\hat{V}$ on $(h_s)_{AB}T$ we need to
write this function in terms of extended spin-network functions.
To that end, just write $h_e=h_s h_{\bar{s}}$ etc., where $\bar{s}$ is 
the non-empty rest of $e$, and apply 
the Clebsh-Gordan decomposition theorem to $h_s\otimes\pi_{j_e}(h_s)$.
The result is given in \cite{14}. Next, apply $\hat{V}$ and obtain a linear
combination of extended spin-network states which we multiply with
$h_s^{-1}$. Upon applying repeatedly again the Clebsh-Gordan decomposition 
and contracting with $[h_\alpha-h_\alpha^{-1}]$ we obtain a gauge invariant
spin-network state which depends on the arc $a$ through spin $1/2$ and
in which $s',s^{\prime\prime}$ carry spin $j_{e'}\pm 1/2,j_{e^{\prime\prime}}
\pm 1/2$ respectively while the spin of $s$ is still $j_e$.\\
So we know how to compute the actions of $\hat{V},\hat{H}^E(N)$ and therefore
of $\hat{K}$. Finally, in order to compute the action of $\hat{T}(N)$
we have to first apply the Clebsh-Gordan decomposition to $h_s T$ and then
evaluate $\hat{K}$ and so forth.\\
Detailed examples of such a computation will be subject to future 
publications \cite{27a}.

\section{The symmetric operator}

This section is devoted to a detailed analysis of a symmetric version of 
the Wheeler-DeWitt operator. The definition of such an operator turns out 
to be a very hard task and the discussion will reveal how tightly the
issues of anomaly-freeness, symmetry and the choice of a regularization 
are interrelated.

\subsection{Motivation}

We argued that the kernel of the non-symmetric operator 
$\hat{H}(N)$, when viewing $0$ as an element of the point spectrum, and 
which consists of 
cylindrical functions on graphs which are such that the tangents of 
edges incident at a vertex are co-planar for each vertex, is too small.
One might think that this kernel 
is incomplete since we stuck with square integrable eigenvectors and that 
it can be enlarged by allowing for more general, distributional solutions 
$\psi\in\Phi'$ to $\hat{H}(N)\psi=0\forall N$. In this case,
as outlined in sections 2 and 6 of \cite{0} we would like to solve the 
Hamiltonian constraint
before the diffeomorphism constraint. We will now see that even so the 
triangulation
prescription made for the non-symmetric operator seems to allow only for 
trivial distributional solutions to the Hamiltonian
constraint.\\ 
The problem already occurs at the level of only the Euclidean
Hamiltonian constraint so let us focus our attention only on $\hat{H}^E(N)$.
Let us try to solve the constraint for graphs with valence higher than 
two. Consider
a function f cylindrical with respect to a graph $\gamma$ and let $v$ be a
non-trivial (in the sense specified above) vertex of $\gamma$ with valence
three to begin with. Writing out $\hat{H}^E_v$ in explicit form we have
\be \label{x1}
-3i\ell_p^2\hat{H}^E_v f=\mbox{tr}(\{h_{\alpha_{12}(\Delta)}-
h_{\alpha_{12}(\Delta)}^{-1})h_{s_3(\Delta)},[h_{s_3(\Delta)},\hat{V}])f
+\mbox{ cyclic.}
\ee
Specifically, let $f:=T_{j_1,j_2,j_3}$ be a spin-network state where the edge
$e_i$
incident at $v$ carries spin $j_i>0$ ($s_i(\Delta)$ is a segment of $e_i$).
It is obvious how the expansion of the right hand side
of (\ref{x1}) in terms of spin-network states looks like : it is a sum 
of up to twelve terms : the first four are defined on the graph $\gamma\cup
a_{12}(\Delta)$
where $a_{12}(\Delta)$ carries spin $1/2$, $s_1(\Delta)$ and $s_2(\Delta)$
carry spin $j_1\pm 1/2$ and $j_2\pm 1/2$ respectively while the rest
of $e_1,e_2$ given by $s_1(\Delta)^{-1}\circ e_1,
s_2(\Delta)^{-1}\circ e_2$ carries still spin $j_1,j_2$ and $e_3$ is unchanged
and carries spin $j_3$. Analogous descriptions hold for the other two
combinations $23,31$. So we see that the original graph got extended.
An ansatz for $\psi$ consisting of an infinite sum of spin-networks 
defined on $\gamma$, that is, $\psi=\sum_{j_1,j_2,j_3}
a(j_1,j_2,j_3)T_{j_1,j_2,j_3}$ does not work for $\psi$ to be in the kernel
which can be seen as follows : First of all, each of the three terms in 
(\ref{x1}) produces a topologically distinct graph so in order for $\psi$
to vanish each of the three infinite sums corresponding to these three
distinct graphs has to vanish separately because spin-network states 
defined on different graphs are orthogonal. Next, notice that the spins 
of the ``top part" of the edges $e_1,e_2,e_3$ are unchanged, therefore 
actually each $\hat{H}^E_v T_{j_1,j_2,j_3}$ has to vanish separately because
spin-network states on the same graph but with different spins are orthogonal.
That means that the values of the coefficients $a(j_1,j_2,j_3)$ are
completely irrelevant.
Finally, each of the twelve terms in the expansion of (\ref{x1}) has to 
vanish separately for the same reason. But one can explictly check that 
the coefficients of that expansion are not all simultaneously vanishing.
So $\psi$ is not in the kernel unless $\psi=0$.\\
The first impulse is that the situation might be improved by
choosing $\psi$ to be diffeomorphism invariant, that is, we take
$\psi:=\sum_{j_1,j_2,j_3} [T_{j_1,j_2,j_3}]$ where the bracket indicates
that we sum over all spin-network states defined on the set of graphs 
defined by the orbit of $\gamma$ under diffeomorphisms \cite{18} but with the
same spins, as in definition \ref{def6}.
However, one readily sees that this does not help either, again,
because of the fact that spin-network states defined on distinct graphs
are orthogonal and because if $\gamma,\phi(\gamma)$ are distinct ($\phi$
some diffeomorphism of $\Sigma$) then $\gamma\cup\Delta(\gamma),
\phi(\gamma)\cup\Delta(\phi(\gamma))$ are still distinct irrespective of
the choice of the assignment $\Delta(\gamma)$. So diffeomorphism invariance
does not help.\\
The second impulse is that then we should make a more general ansatz for
$\psi$ including infinite sums of spin-network states defined on different
graphs not necessarily connected by a diffeomorphism. The simplest guess
is to start with two graphs each of them of the form
$\gamma_{ij}=\gamma\cup\alpha_{ij}(\Delta)$ for two distinct choices of
$(ij)$, say $(12),(23)$ and to hope that the terms coming from appending
$a_{23}(\gamma_{12})$ to $\gamma_{12}$ and vice versa 
cancel each other. But that also fails to be true because in appending
$a_{23}(\gamma_{12})$ the point $a_{23}(\gamma_{12})\cap e_2$ lies
topologically closer to $v$ than $a_{12}(\gamma)\cap e_2$ while in appending
$a_{12}(\gamma_{23})$ the point $a_{12}(\gamma_{23})\cap e_2$ lies
topologically closer to $v$ than $a_{23}(\gamma)\cap e_2$. So the resulting 
graphs are topologically different and the corresponding functions cannot 
cancel each other.\\
Obviously the situation does not improve by considering adding even more
graphs or by increasing the valence of $v$. Finally, also considering the 
full Hamiltonian $\hat{H}(N)$ rather than only $\hat{H}^E(N)$ does not
help because $\hat{T}(N)$ contains two factors of $\hat{K}$ and therefore
introduces even more extraordinary three-valent vertices so that there are
no cancellations with terms coming from $\hat{H}^E(N)$ possible.\\ 
So it seems that $\hat{H}(N)$ does not have a bigger space of solutions
than the one outlined above and we are naturally led again to consider $0$ 
not as an element of the point spectrum but as a point of 
the residual spectrum of $\hat{H}(N)$ (equivalently, as a point of the 
point spectrum of $\hat{H}(N)^\dagger$).

A different factor ordering of the expression of the constraint does not 
help to expand the kernel of $\hat{H}(N)$ because the reason of failure 
to find generalized zero eigenvectors of $\hat{H}(N)$ does not have to do 
with the factor ordering, it has to do with the choice of 
loop-assignment so that it seems that the only 
way out is to modify the triangulation, the only freedom that we have
not exploited yet. \\
It turns out that the requirement of having a symmetric operator,
which is attractive because it removes the quantization ambiguity
of whether to choose $\hat{H}(N)$ or $\hat{H}(N)^\dagger$ as the 
constraint, {\em forces}
us to modify the loop assignment and at the same time enables us to 
enlarge the (distributional) kernel. We will see that the 
obstacle to find a 
symmetric operator is the same as the one that we encountered above : it 
is the fact that the 
repeated action of the Hamiltonian constraint enlarges the graph of a 
cylindrical function without limit.

\subsection{The symmetric Euclidean Operator}

We will prove only those properties of the symmetric operators which are not 
shared by the non-symmetric ones. The reader can convince himself that 
the proofs of cylindrical consistency, diffeomorphism covariance and 
anomaly-freeness as given in the previous paper are entirely unaffected by 
the modifications introduced in the subsequent subsections.

\subsubsection{Symmetry}

We still did not show that, with the symmetric version of definition 
\cite{0},(3.10), 
$\hat{H}^E_\gamma$ qualifies as a projection from $\cal H$ to 
$\mbox{Cyl}_\gamma(\agb)$ of a symmetric operator
$\hat{H}^E$ on $\cal H$. To see the source of the obstruction, observe 
that if $\hat{H}$ is 
any self-consistent operator on $\cal H$ and if $f_\gamma,g_{\gamma'}$
are two cylindrical functions then we have
\ba \label{17}
<g_{\gamma'},\hat{H} f_\gamma>&=&<g_{\gamma'},\hat{H}_\gamma f_\gamma>
=<(\hat{H}_\gamma)^\dagger g_{\gamma'},f_\gamma>\nonumber\\
&=&<\hat{H}^\dagger g_{\gamma'},f_\gamma>=
<(\hat{H}^\dagger)_{\gamma'} g_{\gamma'}, f_\gamma >\;.
\ea
It is important to realize that both adjoint operations involved in 
(\ref{17}) are with respect to $\cal H$
and {\em not} with respect to the completion ${\cal H}_\gamma$ of 
$\mbox{Cyl}_\gamma^3(\agb)$ with respect to the projected measures
$\mu_{0,\gamma}$, see section 2 of \cite{0}.\\
Replacing $\hat{H}$ by $\hat{H}^E$ and using its (anticipated) symmetry as 
well as the one of its projections on $\cal H$
we find that a necessary and sufficient criterion for 
$(\hat{H}^E)^\dagger=\hat{H}^E$ is
\be \label{18}
<g_{\gamma'},\hat{H}^E_\gamma f_\gamma>
=<\hat{H}^E_{\gamma'} g_{\gamma'}, f_\gamma>\;.
\ee
We now will demonstrate that the definition of the triangulation assignment 
given in section 3.1.3 of \cite{0} fails to satisfy this criterion :
In order to see this it is sufficient to check it on a spin-network basis.
So, let $f_\gamma,g_{\gamma'}$ be two spin-network states. Then we see that
$\hat{H}^E_\gamma f_\gamma$ can be written as a finite sum of spin-network
states each of which depends on a common graph $\tilde{\gamma}$ which 
contains $\gamma$ and all the arcs $a_{ij}(\Delta)$ of 
the tetrahedra with basepoint in one of the vertices. 
Notice that by choosing the values of the spin quantum numbers involved 
in $f_\gamma$ large enough we can arrange that the dependence of all these 
spin-networks on all the edges of 
of $\gamma$ and precisely one of the arcs $a_{ij}(\Delta)$ is non-trivial
because of the $\hat{p}_\Delta$ involved in \cite{0},(3.10). Orthogonality 
of the spin-network states therefore implies that the left hand side 
of (\ref{18}) is non-vanishing if and only if $\gamma\subset\gamma'
\subset\tilde{\gamma}$. On the other hand, if indeed $\gamma'=\gamma\cup
\Delta(\gamma)$ where $\Delta(\gamma)$ is one of the tetrahedra assigned 
to $\gamma$ such that $<g_{\gamma'},\hat{H}^E_\gamma(N)f_\gamma>\not=0$
then $\hat{H}^E_{\gamma'}(N)g_{\gamma'}$ can be written as a linear 
combination of spin-network states each of which is {\em bigger} than
$\gamma'$ and therefore $<\hat{H}^E_{\gamma'}(N)g_{\gamma'},f_\gamma>=0$
which contradicts symmetry. The reason why with the loop assignment made 
so far the operator $\hat{H}^E(N)$ is not symmetric comes from the 
requirements 1b) and 1ii) in section 3.1.3 of \cite{0} made for the segments 
$s_I$ of edges $e_I$ and the arcs $a_{IJ}$ assigned to
pairs  of edges $e_I,e_J$ of $\gamma$ incident at a vertex 
$v$ : this requirement basically said that $s_I$ only intersects one 
vertex of $\gamma$, namely $v$, and that $a_{IJ}$ intersects $\gamma$ 
only in its endpoints. Therefore the assignment made for a graph on which
$\hat{H}^E_\gamma(N)f_\gamma$ depends can never coincide with that for 
$\gamma$ itself. \\
One could fix the situation as follows : what needs to be done is to
compute the matrix elements $H^E(N)_{f f'}:=<f,\hat{H}^E(N) f'>$ for any
two cylindrical functions $f,f'$ and then define the matrix elements of
a new symmetric operator $\hat{H}^E(N)_{symm}$ by 
$<f,\hat{H}^E(N)_{symm} f'>:=\frac{1}{2}[H^E(N)_{f f'}
+\overline{H^E(N)_{f' f}}]$.\\ 
While this is what one should do given
the assignment defined in section 3.1.3 of \cite{0} it is a rather indirect 
procedure because we do not know these matrix elements in explicit form.
We prefer to suggest a modification of the assignment and then show that 
(\ref{18}) 
follows. At the moment we are able to do that only at the prize of 
introducing a new structure.
\begin{Definition} \label{def1}
i) A vertex $v$ of a graph $\gamma$ is said to be exceptional provided 
that :\\ 
a) it has at least valence three\\
b)  all edges of $\gamma$ incident at $v$ have co-linear tangents at 
$v$ and precisely two of them, call them $s_1,s_2$, are such that
$s_1\circ s_2$ is an analytic edge\\
c) if $v'$ is any other vertex of $\gamma$ satisfying a) and b) then 
there exists at most one edge of $\gamma$ such that $v,v'$ are its
endpoints.\\
ii) An analytical edge $e$ of a graph $\gamma$ is said to be exceptional 
provided that \\
a) the two vertices $v,v'$ of $\gamma$ corresponding to the endpoints of $e$
are exceptional\\
b) there is a vertex $v_0$ of $\gamma$ and outgoing edges $e_1,e_2$ incident 
at it 
such that $v,v'$ is the endpoint of $e_1,e_2$ distinct from $v_0$\\
c) if the orientation of $e$ is such that it starts at $v$ and ends at 
$v'$ then the tangents of $e,e_1$ at $v$ are parallel and of  
$e,e$ at $v'$ are anti-parallel.
\end{Definition} 
Note that the notion of exceptionality of vertices and edges is 
diffeormorphism invariant and that an exceptional edge can be an
analytical edge. The next definition maps us out of the purely analytical
category.
\begin{Definition} \label{def2}
A smooth exceptional edge $e$ is an edge with all the properties of an
analytical exceptional edge but with the following additional feature :\\
If $v,v'$ are the endpoints of $e$ and $s_1,s_2$ are the edges incident
at $v$ mentioned in definition (\ref{def1}) i),b) such that $s:=s_1\circ 
s_2$ is an analytical edge and likewise if 
$s_1',s_2'$ are incident at $v'$ such that $s':=s_1'\circ s_2'$ is an
analytical edge then $e$ joins $s,s'$ in a $C^\infty$ fashion.
\end{Definition}
Notice that a smooth exceptional edge cannot be analytical : since
all its derivatives coincide with those of $s,s'$ at $v,v'$ it would 
follow from analyticity that $s,s',e$ have coinciding maximal analytical 
extension in contradiction to the fact that $v=e\cap s,v'=e\cap s'$
are only a two points.\\
The idea of how to achieve symmetry is now clear : 
the Hamiltonian constraint defined so far adds 
new edges to a given graph. What one would like to do is to say 
that if $\gamma'$ is a graph which comes from a smaller graph in the sense
that $\gamma'-\gamma$ is a collection of edges
which were added to $\gamma$ by acting repeatedly with the Hamiltonian 
constraint then the action of the Hamiltonian constraint on $\gamma'$ should
coincide with that on $\gamma$. If no such $\gamma$ exists then one 
can choose a loop assignment for $\gamma'$ according to the rules 
described in section 3.1.3 of \cite{0}. The trouble with this strategy is 
that \\
1) it is far from clear that one can construct a consistent loop assignment 
such that for given $\gamma'$ there is at most one $\gamma\subset\gamma'$
such that $\gamma'$ comes from $\gamma$ in the sense just explained 
(so that one would not know how to act with the constraint operator) and\\
2) since $\gamma'$ can just be a given graph and does not necessarily
arise from acting with $\hat{H}^E(N)$ it is intuitively wrong to have 
the ``little edge" $a_{IJ}(\Delta)$ coincide with an edge already 
existing in $\gamma'$ because if one would now make the triangulation 
finer one would need to do that by simultaneously changing the graph 
itself.\\
The following modification of the loop-assignment in section 3.1.3 
of \cite{0} adapted
to the case where the constraint should be a symmetric operator 
circumvents these problems :\\
We keep all points 0),2),4),5). However, 
we introduce the following changes.
\begin{itemize}
\item[6)] {\em Anomaly-Freeness} :\\ 
As we have seen in the main text, a solution to the anomaly-freeness 
condition can be 
given by the following quite simple requirement : each tetrahedron 
$\Delta,\;v(\Delta)\in V(\gamma)$ is subject to the condition that
the loop $\alpha_{ij}(\Delta):=s_i(\Delta)\circ
a_{ij}(\Delta)\circ s_j(\Delta)$ {\em is a kink with vertex at $v$} ! 
That is, the arc $a_{ij}$ joins $s_i,s_j$ in at least a $C^1$ fashion. 
We choose the tangent direction of $a_{ij}$ such that it is parallel to
the one of $s_i$ at $s_i\cap a_{ij}$ and antiparallel to the one of
$s_j$ at $s_i\cap a_{ij}$.
\item[1')] {\em Segments and arcs} :\\
Moreover, to satisfy the symmetry requirement we modify point
1) of section 3.1.3 of \cite{0} as follows :\\
Let $\Gamma$ again be the set of piecewise analytic graphs.
Given $\gamma_0\in\Gamma$ let now the edge $a_{ij}(\Delta)$ be
a {\em smooth} exceptional edge of the graph $\gamma\cap\a_{ij}(\Delta)$
(thus, requirement 6) is met). We keep all the requirements 
of section 3.1.3. of \cite{0} for the 
$s_i(\Delta(\gamma_0)),a_{ij}(\Delta(\gamma_0))$.\\
The image of the $n-th$ power of $\hat{H}^E(N)$ on functions cylindrical 
with respect to
piecewise analytical graphs are now functions on graphs $\gamma_n$ which are 
piecewise analytic after removing precisely $n$ smooth exceptional edges.
The loop assignment for such graphs $\gamma_n$ is then defined 
inductively as follows :\\
i) if $e_I$ is a piecewise analytic edge of $\gamma_n$ necessarily incident 
at a non-exceptional 
vertex $v$ then let $s_I(\gamma_n)$ incident at $v$ be chosen such 
that in case of\\
Situation A : the endpoint of $e_I$ distinct from $v$ is not an endpoint
of a smooth exceptional edge of $\gamma_n$; then apply the rules of section 
3.1.3 of \cite{0} to choose $s_I(\gamma_n)$.\\
Situation B : the endpoint of $e_I$ distinct from $v$ is an endpoint
of a smooth exceptional edge; then choose $s_I(\gamma_n):=e_I$.\\
ii) if $e_I$ is either a piecewise analytic edge of $\gamma_n$ incident 
at an exceptional vertex $v$ or a smooth exceptional edge, necessarily
incident at an exceptional vertex $v$ then 
let $s_I(\gamma_n)$ incident at $v$ be chosen according to the rules of 
section 3.1.3 of \cite{0}.\\
iii) if $e_I,e_J$ are both piecewise analytic edges of $\gamma_n$ 
necessarily incident at a non-exceptional vertex $v$ then there is either
a smooth exceptional edge $a_{IJ}$ connecting the endpoints of $e_I,e_J$
distinct from $v$ or there is not. In the former case we choose 
$a_{IJ}(\gamma_n):=a_{IJ}$, in the latter we apply the rules of section
3.1.3 of \cite{0} to choose $a_{IJ}$ with the addition that $a_{IJ}$ is a 
smooth exceptional edge.\\
iv) if at least one of the two edges of a pair $e_I,e_J$ incident at $v$ 
is an exceptional edge then $v$ is necessarily an exceptional vertex and 
we apply the rules of section 3.1.3 of \cite{0} to choose a smooth 
exceptional edge $a_{IJ}$.\\
It will be shown that the exceptional vertices of $\gamma_n$ do not
contribute to the action of the constraint. It follows that the repeated 
action of the Euclidean Hamiltonian constraint produces functions 
cylindrical with respect to only a finite number of graphs, each of which
has the same unique analytic ``skeleton" obtained by removing its smooth 
exceptional edges. The uniqueness property follows from the fact that the 
exceptional edges are not analytic, they are ``marked" and that was the 
virtue of the construction.\\
Notice that if we have two graphs $\gamma_n,\gamma_n'$ 
which come from the n-th power of $\hat{H}^E(N)$ so that they
have both $n$ smooth exceptional 
edges connecting the same pairs of piecewise analytic edges of their common
skeleton then $\gamma_n,\gamma_n'$ will in general not coincide but they 
will be diffeomorphic. This will be shown in the next point 3').
\item[3')] {\em Diffeomorphism invariant prescription of the position of the
arcs $a_{ij}(\Delta)$} :\\ 
Point 3) of section 3.1.3 of \cite{0} does not quite cover the present 
situation yet because we introduce exceptional edges which in contrast to
section 3.1.3. of \cite{0} {\em always are incident at the same 
exceptional vertex} 
provided they have an endpoint on a piecewise analytic edge of the 
skeleton of the graph under investigation. So given a pair of piecewise 
analytic segments $s_1,s_2$ incident at a non-exceptional vertex $v$ of 
$\gamma$ the requirements of section 3.1.3 of \cite{0} make already sure 
that the 
smooth exceptional arc $a$ connecting $s_1,s_2$ at their endpoints distinct
from $v$ does not intersect any other piecewise analytic segment $s$
incident at $v$. Now, if there are already smooth exceptional arcs $a_1,a_2$
between $s_1,s$ and $s_2,s$ respectively, then in the frame adapted to
$s_1,s_2$ as indicated in section 3.1.3 of \cite{0} we can further specify 
the diffeomorphism in such a way that $a_1,a_2$ do not intersect the part 
of the $x/y$ plane bounded by $s_1,s_2$, except in their endpoints. That 
this is always possible follows from 
the fact that we already found a diffeomorphism adapted to $s_1,s_2$ such 
that $s$ lies either above or below the $x/y$ plane or that it lies 
outside the part of the $x/y$ plane bounded by $s_1,s_2$. Since we can apply
a smooth diffeomorphism to $a_1,a_2$ which preserves the rest of the graph,
the assertion follows. 
\end{itemize}
Since the notion of smooth exceptionality is invariant under analyticity
preserving diffeomorphisms and since we have shown that the assignment
subject to the above modification of our triangulation adapted to a graph
is diffeomorphism covariant, none of the properties proved before in \cite{0}
are ruined.
\begin{Definition}
Consider the range of finite powers of the Euclidean Hamiltonian 
constraint on 
functions cylindrical with respect to graphs in $\Gamma$. These functions 
depend on extended graphs $\gamma$ with an analytic skeleton
$\gamma_0=\gamma-S(\gamma)\in\Gamma$ where $S(\gamma)$ is the set of smooth 
exceptional edges of $\gamma$. We call 
$\Gamma_e$ the set of extended graphs so obtained and $\Gamma_e(\gamma_0)$
is the subset of $\Gamma_e$ consisting of graphs with skeleton
$\gamma_0\in\Gamma_0$.
\end{Definition}
As we have seen, an immediate consequence of this prescription is that an 
(extended) graph $\gamma$ does not grow under the repeated action of the 
Hamiltonian constraint beyond one with a certain finite number of smooth 
exceptional edges. This is in contrast
to the prescription made in section 3.1.3 of \cite{0} and it seems that this 
property 
is forced on us by the requirement of symmetry. The dynamical consequence
of this is a very different structure of the kernel of the constraint
(see next sections).\\ 
The reader may feel uneasy with this prescription because once we have left
the analytic category of graphs we are losing many of the properties 
of the holonomy algebra \cite{7,8,9} and one worries that the quantum 
configuration space $\agb$ is altered. This is because,
if we multiply cylindrical functions defined on finite piecewise 
analytic graphs, the resulting function is a function defined on the union
of the two graphs and the analyticity of the graphs prevents this union
from being an infinite piecewise analytic graph so that the cylindrical 
functions form an algebra. Now if we define the extended graphs to be 
those which have a finite piecewise analytic skeleton after removing
a finite number of smooth exceptional edges then it is easy to see that 
cylindrical functions on extended graphs do not form an algebra.
However, we do not want to do that : we view functions cylindrical with 
respect to extended analytical graphs as states in the Hilbert space
$\cal H$ and as such we cannot multiply them. We still use only
functions which are defined on $\Gamma_0$ to define the spectrum
$\agb$. The only source of non-linearity is the inner product. Now, when 
computing the inner product between functions
cylindrical with respect to extended graphs we make use of the fact that
in order that the inner product be non-vanishing, their skeletons must 
coincide and if so, then the smooth exceptional edges are finite in 
number and mutually non-intersecting and therefore weakly independent 
\cite{7} so that the inner product can easily be computed. This is 
different from inner products between functions cylindrical with 
respect to general smooth graphs and requires more sophisticated 
techniques as for instance in \cite{BaSa}.\\
We confess, however, that a technique that prevents us from introducing
the notion of a smooth exceptional edge and thus leaving the analytical 
category would be strongly preferred. Unfortunately, at the moment we do 
not have such a technique at our disposal.\\

The assertion that with this assignment the family of projections 
$(\hat{H}^E_\gamma(N))$ qualifies as a symmetric operator now 
follows from the following lemma.
\begin{Lemma} \label{la1}
Let $\gamma$ be a piecewise analytic graph, let 
$\gamma'\in\Gamma_e(\gamma)$ and let $f$ be any cylindrical 
function thereon. Then $\hat{H}^E_{\gamma'}f=\hat{H}^E_\gamma f$.
\end{Lemma}
Proof :\\
By construction we just need to check that the edges of 
$V(\gamma')-V(\gamma)$ do not contribute.\\
Consider a function f cylindrical with respect to $\gamma'$ and let
$v\in V(\gamma')-V(\gamma)$. Consider the term $\hat{h}^E_\Delta f$
for any $\Delta$ such that $v(\Delta)=v$. Writing out the 
anti-commutator involved in this term we get two terms. The first is
proportional to
\[ \epsilon^{ijk}\mbox{tr}(h_{\alpha_{ij}(\Delta)}
h_{s_k(\Delta)}[h_{s_k(\Delta)}^{-1},\hat{V}])f=
-\epsilon^{ijk}\mbox{tr}(\{h_{\alpha_{ij}(\Delta)}
h_{s_k(\Delta)}\hat{V} h_{s_k(\Delta)}^{-1})f
\]
where we have used the $SU(2)$ Mandelstam identity $\mbox{tr}(h_\alpha)
=\mbox{tr}(h_\alpha^{-1})$ to simplify the commutator. The volume operator
acts on the cylindrical function $h_{s_k(\Delta)}^{-1}f$ which depends on 
the graph $\gamma'$. Accordingly we can write out $\hat{V}=\sum_{v'\in 
V(\gamma')}\hat{V}_{v'}$ where $\hat{V}_{v'}$ acts only on those edges 
of $\gamma'$ which are incident at $v'$, using the notation of 
\cite{0},(2.8). Take any $v'\not=v$, then the corresponding contribution in 
the 
above expression vanishes because then $h_{s_k(\Delta)}^{-1}$ commutes with
$\hat{V}_{v'}$ and using the $SU(2)$ Mandelstam identity again we see 
that the result is zero. Now if $v'=v$ then the contribution vanishes 
anyway because $v$ is by construction a vertex such that all edges 
incident at it have mutually colinear tangents.\\
Let us now turn to the second term. It is proportional to     
\[ \epsilon^{ijk}\mbox{tr}(h_{s_k(\Delta)}[h_{s_k(\Delta)}^{-1},\hat{V}]
h_{\alpha_{ij}(\Delta)})f=
-\epsilon^{ijk}\mbox{tr}(h_{s_k(\Delta)}\hat{V}h_{s_k(\Delta)}^{-1}
h_{\alpha_{ij}(\Delta)})f
\]
where again use was made of the Mandelstam identity.
The volume operator now acts on the cylindrical function 
$h_{s_k(\Delta)}^{-1}h_{\alpha_{ij}(\Delta)}f$ which depends on the graph
$\gamma'\cup\Delta$ and accordingly the volume operator is now a sum of 
terms $\hat{V}_{v'}$ where $v'$ runs through the vertices of $\gamma'$
and the vertices $v_i(\Delta),i=1,2,3$ of $\Delta$ distinct from 
$v(\Delta)=v$. The only difference to the previous situation is related 
to the additional vertices $v_i(\Delta)$. But these have the same 
property as $v$, namely all incident edges have colinear tangents.
Therefore this contribution vanishes as well.\\
We conclude that all the vertices of $V(\gamma')-V(\gamma)$ are ignored 
by the Hamiltonian constraint and the assertion 
follows now from the cylindrical consistency of the volume operator.\\
$\Box$\\
We notice that if we replace $\hat{h}^E_\Delta$ by $\hat{H}^E_\Delta$
then we find by the same argument (all we used is that the volume 
operator vanishes at vertices which are such that all edges have 
incident tangents) that $\hat{H}^E_{\gamma'}f$ and
$\hat{H}^E_\gamma f$ are diffeomorphic for each $\gamma'\subset
\gamma\cup_{v\in V(\gamma)}\cup_{v(\Delta)=v} \Delta$.\\ 
Using exactly the same arguments as in Lemma \ref{la1} we derive the 
following.
\begin{Corollary} \label{col1}
With the same notation as in \cite{0},(2.8) we have 
\[ 
\hat{h}^E_\Delta f=-\frac{1}{3i\ell_p^2}
\epsilon^{ijk}\mbox{tr}(\{h_{\alpha_{ij}(\Delta)},
h_{s_k(\Delta)}[h_{s_k(\Delta)}^{-1},\hat{V}_{v(\Delta)}]\})f\;.
\]
\end{Corollary}
For $\hat{H}^E_\Delta$ a similar formula holds (just drop the anticommutator
and multiply by 2).
\begin{Theorem} \label{th5}
The system of symmetric projections $\hat{H}^E_\gamma(N)$ defined on 
$D_\gamma$ in \cite{0},(3.10) defines a symmetric operator $\hat{H}^E(N)$ on 
$D$. \end{Theorem} 
Proof :\\
First of all, since the symmetric version of \cite{0},(3.10) involves two 
projectors $\hat{p}_\Delta$, one before and one after acting with
$\hat{h}^E_\Delta$, it follows that either $f_\gamma$ depends non-trivially
on all three $s_i(\Delta)$ before and after acting with $\hat{h}^E_\Delta$
or $\hat{h}^E_\Delta f_\gamma=0$. Thus the right hand side of (\ref{18}) is 
non-vanishing if and only if $\gamma\subset\gamma'\in\Gamma_e(\gamma)$.
By Lemma \ref{la1} we may replace $\hat{H}^E_{\gamma'}$ by 
$\hat{H}^E_\gamma$ on the left hand side of
(\ref{18}). The assertion follows now from  the symmetry of the
operators $\hat{H}^E(N)_\gamma$. \\
$\Box$\\
Before closing this section we would like to point out the following
observation :\\
The requirement that
the loops assigned to an (extended) graph are kinks seems to be forced on
us by anomaly-freeness (compare Theorem \cite{0},3.1).
But as we saw in the proof of the lemma, the kink property
was also important to prove symmetry. So it seems that symmetry and 
anomaly-freeness are tightly knit with each other. 
We see explicitly that the choice of a triangulation adapted to a graph 
is not only a kinematical element of the quantum theory. It is also a very 
dynamical ingredient.

\subsubsection{Self-adjointness}

In the sequel an exceptional edge is a smooth exceptional edge and it is 
understood that in all cylindrical constructions $\Gamma$ is replaced by
$\Gamma_e$.\\
We have shown that $\hat{H}^E_\gamma$ is a symmetric operator on $\cal H$
with dense domain $D_\gamma:=\mbox{Cyl}_\gamma^3(\agb)$, the thrice 
differentiable functions cylindrical with respect to the graph $\gamma$. 
For $\gamma\subset\gamma'$ let $p_{\gamma\gamma'}^\star\;:\; 
\mbox{Cyl}_\gamma(\agb)\to\mbox{Cyl}_{\gamma'}(\agb)$ be the pull-back
of functions from smaller to bigger graphs. The object
$\hat{H}^E$ is defined by the family of projections $(\hat{H}^E_\gamma,
D_\gamma)_\gamma$ and in order to qualify as an operator defined on 
$D=\mbox{Cyl}^3(\agb)$ it needs to be cylindrically consistent up to
a diffeomorphism, that is, 
$(\hat{H}^E_{\gamma'}){|\gamma}
,\hat{H}^E_\gamma$ are diffeomorphic and 
$p_{\gamma\gamma'}^\star D_\gamma\subset D_{\gamma'}$. In the main 
text we have shown that this is indeed the case.\\
If we could show that each $\hat{H}^E_\gamma[N]$ is an essentially 
self-adjoint operator on ${\cal H}_\gamma$ with core $D_\gamma$ then we 
could conclude immediately from theorems proved in \cite{11} that the 
self-adjoint extensions are cylindrically consistent. There are two
reasons why those theorems are inapplicable in our case :\\
1) The range of $\hat{H}^E_\gamma$ on $D_\gamma$ does not lie in 
${\cal H}_\gamma$ since $\hat{H}^E_\gamma$ enlarges the graph by the arcs.\\
2) While one could try to circumvent that problem by considering 
$\hat{H}^E_\gamma$
as an operator on ${\cal D}_{\tilde{\gamma}}$ where $\tilde{\gamma}$ is a 
graph on which all cylindrical functions in the range of powers of
$\hat{H}^E(N)$ on ${\cal D}_\gamma$ depend, according to lemma \ref{la1},
(and it turns out that it then is symmetric on $D_{\tilde{\gamma}}$) we 
simply do not know whether that operator is essentially self-adjoint
on ${\cal D}_{\tilde{\gamma}}$.\\
The way out is to work directly on the full Hilbert space $\cal H$
which is the completion with respect to the obvious inner product
of the space $\cup_\gamma {\cal H}_\gamma$ (again we did not display
identifications due to cylindrical equivalence).\\
To see that each $\hat{H}^E$ has self-adjoint extensions we use a
theorem due to von Neumann (\cite{24}, p. 143).
\begin{Definition}
An antilinear map $\hat{k}\;:\;{\cal H}\to {\cal H}$ is called a conjugation
if it is norm-preserving and $\hat{k}^2=\mbox{id}_{{\cal H}}$.
\end{Definition}
\begin{Theorem}[von Neumann's theorem]
Let $\hat{H}$ be a symmetric operator on a Hilbert space with dense 
domain $D$ and suppose that there exists a conjugation $\hat{k}$
satisfying the following two properties :\\
1) $\hat{k} D\subset D$ preserves the domain and \\
2) $\hat{k}\hat{H}=\hat{H}\hat{k}$ on $D$, that is, $\hat{H}$ commutes with
the conjugation.\\
Then $\hat{H}$ has self-adjoint extensions.
\end{Theorem}
The proof follows from the fact that the assumptions imply that $\hat{H}$
has equal deficiency indices.\\
To apply this theorem to our case we begin by noticing that 
$D=\mbox{Cyl}^3(\agb)=\cup_\gamma \mbox{Cyl}^3_\gamma(\agb)$
is a dense domain for $\hat{H}^E$ and that it is spanned by spin-network
states. 
But these states can be expanded, with real coefficients, into
traces of the holonomy around closed loops (that is, Wilson loop functionals)
and it is a peculiarity of $SU(2)$ that the latter are real valued.
The explicit form of \cite{0},(3.10) implies then that the result of applying
$\hat{H}^E$ will be a sum of spin-network states with {\em purely imaginary}
coefficients, that is, the operator $\hat{H}^E$ is imaginary-valued, its
matrix elements are purely imaginary and anti-symmetric in a basis of real 
valued functions like the spin-network basis. Therefore, it is not enough 
to choose $\hat{k}$ to be just complex conjugation.\\
Given an extended graph $\gamma$, consider its skeleton $\gamma_0$.
Recalling the definition of a smooth exceptional edge, by 
inspection of \cite{0},(3.10) each of the six terms involved in 
$\hat{h}^E_\Delta$ depends precisely one one smooth exceptional
edge $a_{ij}(\Delta)$. Therefore, given a spin-network state $f$
cylindrical with respect to $\gamma$, if we expand 
the state $\hat{H}^E_\Delta f$ as a linear combination of
spin-network states, then each of those states depends on a graph 
$\gamma'$ such that the spin associated with precisely one of the
smooth exceptional edges assigned to $\gamma_0$ has changed in $\gamma'$ by 
$\pm \hbar/2$ as 
compared to $\gamma$ (to see this, consider $f$ as a state on $\gamma'$).
We are going to exploit precisely this fact to construct an appropriate
conjugation.
\begin{Theorem}
The operator $\hat{H}^E(N)$, densely defined on $\mbox{Cyl}^3(\agb)$,
possesses self-adjoint extensions.
\end{Theorem}
Proof :\\
Denote the exceptional edges of a graph $\gamma$ by $E_0(\gamma)$.
Let $e\in E_0(\gamma)$ and $Y^i_e:=X^i(h_e)$, where $X^i$ is the 
right invariant vector field on $SU(2)$, and construct their Laplacians
$\Delta_e:=\mbox{tr}(Y_e Y_e)$ which are negative definite operators on 
$SU(2)$ with usual eigenvalues $-j(j+1)$ on eigenfunctions with spin $j$,
in particular on spin-network states.
We construct a positive definite spin operator 
$\hat{J}_e:=\sqrt{\frac{1}{4}-\Delta_e}-\frac{1}{2}$ with eigenvalues 
$j$. Finally we set 
\be \label{19}
\hat{P}_\gamma:=\prod_{e\in E_0(\gamma)} e^{2 i\pi\hat{J}_e}\mbox{ and }
\hat{k}_\gamma:=\hat{P}_\gamma\hat{c}
\ee
where $\hat{c}$ is the operator of complex conjugation. \\
Obviously each $\hat{P}_\gamma$ has a domain $\mbox{Cyl}_\gamma^2(\agb)$,
dense in ${\cal H}_\gamma$ and is a bounded (by $1$) and symmetric operator
thereon (on a spin-network state it corresponds to multiplying the state
by $\pm 1$). $\hat{P}_\gamma$ is even an essentially self-adjoint operator
on ${\cal H}_\gamma$ with core $\mbox{Cyl}_\gamma^2(\agb)$ : To see this 
we check the basic criterion of essential self-adjointness. 
We need to show that $\hat{P}_\gamma\pm i\mbox{id}_{{\cal H}_\gamma}$ has 
dense range and it will be enough to show that each spin-network $f$ 
state on $\gamma$ is in the range of that operator when evaluated on its 
domain. But $[\hat{P}_\gamma\pm i\mbox{id}_{{\cal H}_\gamma}]T=[\pm 1\pm i]T$
so $T$ is reproduced up to a never vanishing multiplicative factor. That
proves that $(\hat{P}_\gamma,\mbox{Cyl}_\gamma^2(\agb))$ is essentially 
self-adjoint.\\
Let us check that the family $(\hat{P}_\gamma,
\mbox{Cyl}_\gamma^2(\agb))_{\gamma\i\Gamma_e})$
defines an essentially self-adjoint operator $\hat{P}$ on 
$\cal H$ with dense domain $\mbox{Cyl}^2(\agb)$. The condition 
$p_{\gamma\gamma'}\mbox{Cyl}_\gamma^2(\agb)\subset\mbox{Cyl}_{\gamma'}^2(\agb)$
is certainly satisfied for each $\gamma\subset\gamma'$. Since each 
$f\in\mbox{Cyl}_\gamma^2(\agb)$ can be expressed in terms of spin-network
states $T$ it is sufficient to check cylindrical consistency on those
functions. But if $\gamma$ is lacking an exceptional edge $e$ as compared 
to $\gamma'$ then $\hat{J}_e T=0$ proving cylindrical consistency.
This shows that the closure of $\hat{P}$ is even a self-adjoint operator on 
$\cal H$
since it was shown in (\cite{11}) that each consistent and essentially 
self-adjoint family is such that the family of self-adjoint extensions is
cylindrically consistent.\\
Finally it is easy to see that $\hat{P}$ is a linear, norm-preserving 
operator on $\cal H$ upon checking in a spin-network base.
Moreover, $\hat{P}^2=\mbox{id}_{\cal H}$ which follows from 
$\hat{P} T=\pm T$ for any spin-network state, that is, $\hat{P}$ acts 
like a parity operator on exceptional edges.\\
Finally, it follows that $\hat{k}:=\hat{P}\hat{c}$ is a conjugation
on $\cal H$ which follows from the fact that the phase shift of 
a spin-network state $T$ induced by $\hat{P}$ is real and from the fact that
$\hat{P}$ is linear so that $\hat{k}$ is anti-linear.\\
We are now ready to see that $\hat{H}^E[N]$ commutes with $\hat{k}$.
Consider a term 
$\hat{H}^E_{ijk,\Delta}:=-\frac{1}{3i\ell_p^2}
\mbox{tr}(\{h_{\alpha_{ij}(\Delta)},
h_{s_k(\Delta)}[h_{s_k(\Delta)}^{-1},\hat{V}_v]\})$
where $e:=a_{ij}(\Delta)$ in $\alpha_{ij}(\Delta)$ is a smooth exceptional 
edge. Let $T$ be a spin-network state with spin $j_e$ associated with $e$.
Then $\hat{H}^E_{ijk,\Delta}T=f_+ + f_-$ where $f_\pm$ are sums of 
spin-network states such that they depend on $e$ through $j_e^\pm=j_e\pm 
1/2$ while the spins of all other exceptional edges are unchanged (this is 
because the other edges $s_i(\Delta)$ involved in $\hat{H}^E_{ijk,\Delta}$ 
are non-exceptional edges). It follows easily that 
$\hat{P}f_\pm=e^{2i\pi(j_e\pm 1/2)}\prod_{e'\in E_0(\gamma)-e}
e^{2i\pi j_{e'}} f_\pm=-pf_\pm$ where $p$ is defined by
$\hat{P}T=pT$. Thus 
\[ \hat{P}\hat{H}^E_{ijk,\Delta}T=-\hat{H}^E_{ijk,\Delta}\hat{P}T\]
and so $[\hat{k},\hat{H}^E[N]]f=0$ for any $f\in D$ due to the 
factor of $i$ involved in $\hat{H}^E_{ijk,\Delta}$.\\
$\Box$\\
This proves only existence, not uniqueness, of self-adjoint extensions for 
$\hat{H}^E[N]$.
We do not know how many extension there are and how to select one
in case there are several. We conjecture, that $\hat{H}^E[N]$
is even essentially self-adjoint in which case that extension would be 
unique and concisely described by the theorems in \cite{11}. A proof
for that conjecture is missing, however, at the present stage.

\subsection{The symmetric Lorentzian operator}

Again we will only discuss the points of departure between the symmetric and
non-symmetric operators. It is understood that the triangulation as modified
in the previous section is applied to the present section as well. Also,
as discussed in the main text, without changing formula \cite{0},(4.1),
$\hat{K}$ is now automatically a symmetric operator and it has self-adjoint
extensions.

\subsubsection{Symmetry and cylindrical consistency}

It turns out that if we choose the ordering
\be \label{28}
\hat{t}_\Delta:=-\frac{64}{3 (i\ell_p^2)^3}\epsilon^{ijk}
\mbox{tr}(h_{s_i(\Delta)}[h_{s_i(\Delta)}^{-1},K_v]
h_{s_j(\Delta)}[h_{s_j(\Delta)}^{-1},V_v] h_{s_k(\Delta)}
\{h_{s_k(\Delta)}^{-1},K_v\})
\ee
and use that $\hat{V},\hat{K}$ are symmetric operators, as well as the 
$SU(2)$ reality conditions, and use this operator  
in \cite{0}(5.3) then the operator is already symmetric with domain 
$\mbox{Cyl}^3(\agb)$
on $\cal H$. Therefore we replace \cite{0}(5.3) by 
\be \label{28a}
\hat{t}_\gamma[N]:=\sum_{v\in V(\gamma)}\frac{N_v}{E(v)} \hat{t}_v,\;
\hat{t}_v:=\sum_{v(\Delta)=v} \hat{t}_\Delta\;.
\ee
In complete analogy with the discussion for $\hat{H}^E$ we now define 
\be \label{29}
\hat{T}_\gamma[N]:=\sum_{v\in V(\gamma)} N_v\sum_{v(\Delta)=v}
\frac{\hat{p}_\Delta}{\sqrt{\hat{E}(v)}}\hat{t}_\Delta
\frac{\hat{p}_\Delta}{\sqrt{\hat{E}(v)}}
\ee
and arrive at a self-consistent family of symmetric operators. To show 
that this family qualifies as the set of graph projections of a symmetric  
operator on $\cal H$ we need an analogue of Lemma \ref{la1}.
\begin{Lemma} \label{la2}
With the same notation as in Lemma \ref{la1} it holds that 
$\hat{T}_{\gamma'}f=\hat{T}_\gamma f$.
\end{Lemma}
Proof :\\
The proof follows immediately from the fact that the volume operator 
vanishes at the vertices of $V(\gamma')-V(\gamma)$ and the explicit 
expression (\ref{28}) along the same line of argument as in Lemma 
\ref{la1}.\\
$\Box$\\
The proof that then Theorem \ref{th5} holds with $\hat{H}^E$ replaced by 
$\hat{T}$ is completely analogous and is omitted.

\subsubsection{Self-Adjointness}

While we could try to invoke von Neumann's theorem again to prove that
self-adjoint extensions of $\hat{T}$ exist, this is insufficient since
self-adjointness does not respect the linear structure of the operator
algebra. Rather, given some self-adjoint extension $D(\hat{H}^E)$
of $\hat{H}^E$, what we
need is an extension of $\hat{T}$ to the same domain $D(\hat{H}^E)$.\\
An obvious approach to prove existence of such an extension is suggested  
by the following theorem \cite{24}.
\begin{Theorem}[Kato-Rellich]
Suppose that $\hat{H}^E$ is self-adjoint on $\cal H$ with domain 
$D(\hat{H}^E)$ and that $\hat{T}$ is symmetric with domain $D(\hat{T})$
such that $D(\hat{H}^E)\subset D(\hat{T})$. 
Furthermore, suppose that there are real numbers $a,b$ such that for all 
$\psi\in D(\hat{H}^E)$ it holds that $||\hat{T}\psi||\le a||\hat{H}^E\psi||
+b||\psi||$ and that the infimum of all possible $a$ (as $b$ varies) 
satisfies $a<1$.
Then $\hat{H}:=\hat{T}-\hat{H}^E$ is self-adjoint on $D(\hat{H}^E)$.
\end{Theorem}
To apply this theorem we therefore need to perform three steps :\\
a) Choose a self-adjoint extension of $\hat{H}^E$,\\
b) Check whether there is a domain of $\hat{T}$ which contains the 
determined domain of $\hat{H}^E$ and \\
c) check whether the bound condition mentioned in the theorem (which
in the mathematics literature is called ``$\hat{T}$ is 
$\hat{H}^E$-bounded with relative bound $<1$") can be satisfied for 
some choice of $b$.\\
Clearly, such an analysis is far from trivial and is beyond the scope of 
the paper. We will get back to this question in a later paper and just
comment on why we can hope to find a relative bound $<1$ : 
A dense domain of $\hat{H}^E$ are the finite linear combinations of 
spin-network states on which also $\hat{T}$ is symmetric so that it is 
plausible that the first condition in the theorem is satisfied. If $N$ 
is the total number of edges of a graph $\gamma$ define $j:=j_1+..+j_N$ 
for a spin network state $\psi$ with spins $j_1,..,j_N$. It follows from
elementary angular momentum algebra that $||\hat{V}\psi||\le j^{3/2}||\psi||$
(here we used the boundedness of the matrix elements of an element of 
$SU(2)$ ).
Moreover, since $h_e$ changes the spin associated with the edge $e$ by
$\pm\hbar/2$, it follows that $||h_e[h_e^{-1},\hat{V}]\psi||\propto
j^{1/2}||\psi||$. We thus expect a behaviour like $||\hat{H}^E_v\psi||
\propto j^{1/2}||\psi||$. Next, recall that 
$\hat{K}\propto[\hat{V},\hat{H}^E[1]]$ so that we find $||\hat{K}\psi||
\propto j||\psi||$ which means that by a similar argument also
$||\hat{T}_v\psi||\propto j^{1/2}||\psi||$. So the large spin behaviour 
of both
$\hat{T},\hat{H}^E$ is comparable and it is conceivable that a relative
bound $a<1$ exists given the fact that in $\hat{T}$ a lot more 
symmetrizations among the edges are taking place.

\subsubsection{Solutions}

The detailed analysis of the kernel of $\hat{H},\hat{H}^E$ will be
left for future publications \cite{27a}. Here we content ourselves with
a qualitative description.

1) The most important property of the symmetric operator is that it does 
not extend
a given analytic graph $\gamma$ beyond graphs contained in $\Gamma_e(\gamma)$
as described in Lemma \ref{la1}. If we work on diffeomorphism invariant states
then there is even a maximal, finite graph $\tilde{\gamma}$ on which 
(diffeomorphic images of) all $\gamma'\in\Gamma_e(\gamma)$ depend.
This implies that we can study the eigenvalue problem on the finite graph
$\tilde{\gamma}$, that is, instead of dealing with $\cal H$ we just have 
to consider its projection ${\cal H}_{\tilde{\gamma}}$ which turns the
spectral analysis into a problem on a Hilbert space with a finite number 
of degrees of freedom. In particular, since we know that all the 
spin-network states on that graph form a complete set of orthonormal states, 
this Hilbert space is separable. \\
In particular, this property is precisely the reason why now an infinite 
series of spin-networks on the graph $\tilde{\gamma}$
has a chance to be annihilated by 
$\hat{H}(N)$ upon choosing the coefficients of that 
expansion appropriately. Such a series is a well-defined element of 
$\Phi'$ and we see that again the action of the Hamiltonian and 
Diffeomorphism constraints on spin-netwoks is nicely split : the Hamiltonian 
constraint acts on $\vec{j},\vec{c}$ and leaves $\gamma$ invariant
while the Diffeomorphism constraint acts on $\tilde{\gamma}$ only. This
separation between labels on which the two constraints effectively act on 
is the deeper reason for the fact that the constraint algebra of the 
symmetric Hamiltonian constraint is effectively Abelian.

2) If we can at least prove existence of self-adjoint extensions then
we can exponentiate the Hamiltonian and compute rigorously defined 
solutions by the group-averaging method \cite{16,17}. 
By the same method we are 
able to find a scalar product on the space of solutions. This is possible 
because the second 
important property of the Hamiltonian constraint is that the operators
corresponding to different vertices commute (in the diffeomorphism 
invariant context) and so far we 
are only able to deal with the group averaging method provided we know the
group that is generated, and a special case of this is when we have a
finite number of Abelian constraints. \\
This goes as follows :\\
On the graph $\tilde{\gamma}$ the Hamiltonian constraint reduces to
$\hat{H}_{\tilde{\gamma}}[N]=\sum_{v\in V(\gamma)}N_v\hat{H}_v$. Suppose
we have found a self-adjoint extension for each of the $\hat{H}_v$
then, by Stone's theorem,  we can exponentiate $\hat{H}_{\tilde{\gamma}}$ 
and obtain a unitary operator
\be \label{37}
\hat{U}[\vec{N}]:=\prod_{v\in V(\gamma)} e^{i N_v \hat{H}_v}
\ee
where $\vec{N}=(N_v)_{v\in V(\gamma)}$. Actually we obtain a unitary 
representation of an $n-$dimensional Abelian group with parameter $\vec{N}$ 
and group 
structure $\hat{U}[\vec{M}]\hat{U}[\vec{M}]=\hat{U}[\vec{M}+\vec{N}]$.
Then the group average proposal says that we take a physical state to be
\be \label{38}
[f]:=\int_S d\mu_H(\{N_v\}_{v\in V(\gamma)})\prod_{v\in V(\gamma)}  e^{i 
N_v\hat{H}_v} f \ee
where $f$ is any function cylindrical with respect to $\tilde{\gamma}$
and $d\mu_H$ is the Haar-measure on the group manifold $S$ 
coordinatized by the $N_v$.
To see that $[f]$ is a solution of the constraint we just verify
that 
\be \label{39}
\hat{U}[M][f]:=\int_S d\mu_H(\{N_v)\}) \hat{U} [M_v+N_v] f =[f]
\ee
since the Haar measure is translation invariant.\\
The inner product induced by the Hamiltonian constraint is given by 
\be \label{40}
<[f],[g]>_{phys}:=<f,[g]>
\ee 
where the inner product on the right hand side is the one on $\cal H$.
This inner product has the feature that whenever we have an
observable on $\cal H$  which commutes with $\hat{H}$ strongly, that 
is, $<f,[\hat{O},\hat{H}] g>=0$ for all $f,g\in\mbox{Cyl}^\infty(\agb)$ 
then it 
projects to an operator $[\hat{O}]$ on ${\cal H}_{phys}$ with preserved 
adjointness relations. Namely 
from $\hat{U}(\vec{N})^{-1}\hat{O}\hat{U}(\vec{N})=\hat{O}$ and 
$<[f],g>=<f,[g]>$ we find upon choosing $[\hat{O}][f]:=[\hat{O}f]$ that 
$[\hat{O}]^\dagger=[\hat{O}^\dagger]$.\\
All these concepts are explained in more detail in \cite{16,17,18}.

\subsection{Wick rotation transform}

As explained in \cite{4} (see also \cite{5}) one has also another option
to define the Wheeler-DeWitt constraint operator provided that the 
generator of the Wick rotation transform is self-adjoint.
But that we checked to be the case and we can proceed and repeat the 
main argument.\\
One can show that there is a classical generator, called the complexifier
in \cite{4}, of the canonical transformation 
$(A=\Gamma+K,E)\to(A=\Gamma-iK,iE)$ and it is just given by $C=(\pi/2) K$.
Then one can show that up to a term proportional to the Gauss constraint
it holds that 
\be \label{32}
H=p(H^E+\{H^E,-iC\}+\frac{1}{2}\{\{H^E,-iC\},-iC\}
+\frac{1}{3!}\{\{\{H^E,-iC\},-iC\},-iC\}+..)
\ee
where $p$ is a phase depending on how we take the square root of
$i^3$, because effectively the real connection $A$ gets replaced by the
complex Ashtekar connection. This expression motivates to just {\em define}
\be \label{33}
\hat{H}=p(H^E+[H^E,-\hat{C}/\hbar]+\frac{1}{2}[[H^E,-\hat{C}/\hbar],-\hat{C}/\hbar]
+\frac{1}{3!}[[[H^E,-\hat{C}/\hbar],-\hat{C}/\hbar],-\hat{C}/\hbar]+..)
\ee
or, upon defining the Wick rotation operator
\be \label{34}
\hat{W}:=e^{-\hat{C}/\hbar}
\ee
we have
\be \label{35}
\hat{H}=p\hat{W}^{-1}\hat{H}^E\hat{W} \;.
\ee
There are three obvious problems :\\
1) Although $\hat{C},\hat{H}^E$ were shown to possess self-adjoint 
extensions, it is unclear whether they possess self-adjoint 
extensions to a common domain (in which case we would have a chance  
that (\ref{35}) makes sense as far as domain questions are concerned).\\
2) The operator $\hat{C}$ is far from being positive definite, therefore
$\hat{C}$ is not the generator of a contraction semigroup given formally by
$\hat{W}_t:=\exp(-t\hat{C}/\hbar)\;,t>0$ and it is unclear whether 
$\hat{W}$ can be defined at all on a dense domain of $\cal H$. One possible
approach would be to restrict the Hilbert space to the ``positive frequency
subspace" where $\hat{C}$ is positive definite (indeed, $1/\ell_p^3 
\hat{C}$ has the dimension of a frequency), however, that could mean that 
we alter the reality conditions. \\
3) Whenever $\hat{W}$ can be defined, it is going to be a symmetric operator.
But then $\hat{H}$ will not even be symmetric and again group averaging 
methods cannot be immediately applied.\\
A way to resolve these problems is suggested by recalling a theorem due to 
Nelson \cite{24}.
\begin{Definition}
Let $\hat{C}$ be an operator on $\cal H$. Then $C^\infty(\hat{C}):=
\bigcap_{n=0}^\infty D(\hat{C}^n)$ is called the set of smooth vectors
for $\hat{C}$ and a vector $\psi\in C^\infty(\hat{C})$ is called
analytic if there exists $t>0$ such that 
\[ \sum_{n=0}^\infty\frac{||\hat{C}^n\psi||}{n!} 
(\frac{t}{\hbar})^n<\infty\;. \] \end{Definition}
\begin{Theorem}[Nelson's analytic vector theorem]
A closed symmetric operator $\hat{C}$ is self-adjoint if and only if 
its domain $D(\hat{C})$ contains a dense set of analytic vectors.
\end{Theorem}
We have shown already that $\hat{C}$ (actually its closure) has a 
self-adjoint extension. Therefore it follows from Nelson's theorem that 
there exists a dense set of analytic vectors for $\hat{C}$ on which
$\hat{W}_t$ actually {\em does} converge in norm for some $t>0$. The 
question then remains if we can choose $t=1$.\\

On the other hand, even if we can choose $t=1$, we are actually 
interested in solving the quantum 
constraint $\hat{H}\psi=0$ and we would like to do that by setting  
$\psi=\hat{W}^{-1}\psi^E$ where $\hat{H}^E\psi^E=0$ is typically a 
{\em distributional} solution of the Euclidean Hamiltonian constraint.
So how can we even hope to solve the Lorentzian constraint by this method ?
The answer is the following : $\psi^E$ is an infinite sum of $L_2$
vectors (which does not converge in $\cal H$ but in $\Phi'$).
Since the set of analytic vectors is dense in $\cal H$, each of these
$L_2$ vectors can be written
as a (infinite) sum of analytic vectors for $\hat{C}$ which converges in 
$\cal H$. In summary we can write $\psi^E$ in terms of analytic vectors 
for $\hat{C}$ and we can apply $\hat{W}$
to each of them separately. Since the result of this is a series of 
$L_2(\agb,d\mu_0)$ vectors we can hope that it makes sense as a 
distribution again, provided we can choose $t=1$ in Nelson's theorem.\\
If it turns out that we cannot choose $t=1$ to define $\hat{W}$ 
or even if it does, that $\hat{W}\psi^E\not\in\Phi'$ then
we may be forced to adopt still another strategy which consists in going
to a holomorphic representation \cite{4}. The point of this is the 
following : one maps a cylindrical function $f$ by $\hat{W}$ and 
then analytically
continues it. This analytic continuation is done for each term 
$\hat{C}^n\psi$ (which is a cylindrical function again and so has a 
well-defined analytic continuation) separately. While the sum of these
terms may not make any sense as an element of $\cal H$ before analytically
continuing it, after analytic continuation it may make sense as a 
distribution on a 
dense subspace space of a Hilbert space of functions of
complex-valued connections upon choosing a measure thereon which 
has the necessary stronger fall-off property. In order to satisfy the
correct reality conditions this measure needs to be chosen in such a way
that $\hat{U}:=\hat{a}\hat{W}$ (where $\hat{a}$ means analytic continuation)
is unitary (see \cite{4} for a more detailed discussion).\\
This could resolve issues 1) and 2) but not 3). One might think that the
part of the algebraic quantization programme that concerns the group 
averaging method is inapplicable because of that. However, while 
we cannot define the unitary evolution of $\hat{H}$ immediately by
exponentiating it since it is not self-adjoint, we can define the unitary 
evolution of $\hat{H}^E$
and then just define $\exp(it\hat{H}):=\hat{W}^{-1}\exp(it\hat{H}^E)\hat{W}$.
The operator $\exp(it\hat{H}^E)$ can then be used to 
define the physical inner product by the group averaging method.\\
The task to answer these questions will be left to future investigations.
As it should be clear, to settle these mathematical issues it is again of 
utmost importance to gain maximum control over the spectrum of the volume 
operator \cite{ThieVol}.

\section{Conclusions}

Let us now summarize what can be said qualitatively about the action of the
Wheeler-DeWitt constraint operator as defined in these two papers.
\begin{itemize}
\item[1)] {\em Action of $\hat{H}^E$} :\\
Spin-network states on a fixed graph are labelled by the spin 
quantum
numbers $j_I$ associated with the edges of the graph and a contraction 
matrix which turns the associated tensor product of irreducible 
representations into a gauge invariant function. Consider first the operator
$\hat{H}^E$. Qualitatively, the part $h_{s_k}[h_{s_k}^{-1},\hat{V}]$ does 
not change the quantum numbers $j_I$ at all, it changes the contraction 
matrix. The part $h_{\alpha_{IJ}}^{\pm 1}$ on the other hand changes the
spins $j_I,j_J$ by $\pm 1/2$. For instance, on a trivalent graph where 
for given $j_1,j_2,j_3$ the contraction matrix is given uniquely, one can
show that the action of the Euclidean Hamiltonian constraint looks like this
\be \label{36}
\hat{H}^E_v T_{j_1,j_2,j_3}=i\sum_{\mu,\nu=\pm 
1/2} c_3(\mu,\nu;j_1,j_2,j_3) T'_{j_1+\mu,j_2+\nu,j_3}+\mbox{cyclic}
\ee
for certain real-valued functions $c_I$ of $j_1,j_2,j_3$ 
and $T_{j_1,j_2,j_3}$ is a spin-network function corresponding to the
spins $j_I$ associated with the edges meeting at $v$ and it is understood 
that the graph with respect to which $T'$ is cylindrical contains one of the 
arcs $a_{IJ}$.\\ 
One sees that the action of $\hat{H}^E$ can be visualized 
as \\ {\bf the annihilation ($\mu=\nu=-1/2$), creation ($\mu=\nu=+1/2$)
and re-routing ($\mu\nu=-1/4$) of spin associated with the graph in units 
of $\Delta j=\pm 1/2$}. This picture is insensitive of whether we are 
dealing with the symmetric or non-symmetric version of the constraint.
In other words, the picture we have is quite similar to the one we have
in Quantum Electrodynamics (QED) : the Hamiltonian of QED is an infinite
sum of uncoupled harmonic oscillators, two for each mode (momentum 
$\vec{k}$). A cylindrical function for QED is a state with a finite 
number $n_I$ of photons of momentum $k_I$ and polarization $p_I$.
On such a cylindrical function the QED Hamiltonian reduces to a 
finite number of harmonic oscillator Hamiltonians each of which is a 
polynomial of annihilation and creation operators which act by 
annihilating and creating the number of photons for the given mode and 
polarization in units of $\Delta n=\pm 1$. The two objects that correspond
to each other in the two theories are first a) the continuous labels 
$\gamma=\vec{e}$ (the edges)
and $\vec{k},\vec{p}$ and secondly b) the discrete quantum numbers or 
occupation numbers $\vec{j}$ and $\vec{n}$.\\
The analogy fails in the respect that we cannot associate elementary
particles (we do not have gravitons, the analogon of photons) with the 
elementary excitations of the gravitational field. What is excited 
are lines of force and
the continuous information that they carry is position rather than momentum.
Thus this Fock representation is based on position rather than momentum.
\item[2)] {\em Action of $\hat{T}$}\\
Let us now consider the operator $\hat{T}$. Since 
$\hat{K}\propto
[\hat{V},\hat{H}^E]$ it follows from the fact that $\hat{V}$ does not alter
representations that also $\hat{K}$ acts by annihilation, creation and
rerouting of spin by $\Delta j=\pm1/2$. Also, it is clear that 
$h_s[h_s^{-1},\hat{K}]$ does not modify the qualitative behaviour of 
$\hat{K}$. It follows then that $\hat{T}$ changes the spin of one edge by 
$\Delta j=-1,-1/2,0,1/2,1$ because there are two factors of $\hat{K}$
involved and the various terms can act on different edges or the same again.
Therefore, the behaviour of $\hat{H}$ and $\hat{H}^E$ are roughly the
same, just the numerical coefficients are different, in principle we can
describe the Wheeler-DeWitt operator as a low order polynomial of degree
two in the creation and annihilation operators associated with the 
spin of the edges. The computation of the precise coefficients of this
polynomial is a tedious but straightforward task. 
In particular, even for the symmetric operator, it seems that the spectrum can
be computed either exactly or with a high degree of precision and that
the self-adjoint extensions can be obtained by direct methods.
\item[3)] {\em ADM energy is diagonal}\\
The analogy with the Fock representation of QED is further enhanced by 
noticing that the ADM-Hamiltonian is diagonal on certain linear 
combinations of spin-network states on one and the same graph, just like 
the QED Hamiltonian which is diagonal on the photon Fock states. So the
ADM-Hamiltonian is essentially an occupation number operator.\\
To see this recall 
that $E_{ADM}=\lim_{r\to\infty}\int_{S_r}dS_a(q_{ab,b}-q_{bb,a})$ where 
$S_r$ is a one-parameter family of two-dimensional surfaces with the
topology of $S^2$ and $r$ is the radius of the sphere as measured by 
a fixed asymptotic flat background metric. Now it follows immediately
from $e_a\propto\{A_a,V\}$ that $q_{ab}$ when integrated over a 
two-dimensional surface has the chance to have a well-defined quantization
and that turns out to be correct \cite{29}. Again, the eigenvalues of 
$\hat{E}_{ADM}$ are certain algebraic function of the spins $\vec{j}$.
This fact motivates to call the spin-network representation 
$|\gamma,\vec{j},\vec{c}>$, defined abstractly by 
$<[A]|\gamma,\vec{j},\vec{c}>=T_{\gamma,\vec{j},\vec{c}}(A)$, $[A]$ the 
gauge equivalence class of $A$, where as usual
$<[A'],[A]>:=\delta_{\mu_0}([A],[A'])$, the ``non-linear Fock-representation"
for the string-like excitations of the gravitational field.\\
All these facts motivate to call the dynamical theory obtained 
``Quantum Spin Dynamics (QSD)" as opposed to ``Quantum Geometrodynamics" or
QED.
\item[4)] {\em Final Comments in order} :\\
$\bullet$ Both, the non-symmetric \cite{0},(5.5) and the symmetric 
(\ref{33}) version are quantizations of the
Wheeler-DeWitt constraint for Lorentzian, four-dimensional quantum gravity
in the continuum which are well-defined on the whole Hilbert space $\cal H$.
In that respect they differ considerably from the operator defined in 
\cite{28} which a) is given on a lattice rather than in the 
continuum, b) is a discretization of the rescaled form of the 
Wheeler-DeWitt constraint with density weight two which 
is possible only on a lattice without capturing the singularities that one
will ultimately encounter in any suitable continuum limit and c) is singular
on a huge subspace of the lattice Hilbert space in any ordering and 
therefore is not even densely defined.\\
$\bullet$ Our Euclidean Hamiltonian constraint operator \cite{0},(3.10) also 
is {\em entirely different} from those proposed in \cite{12,13} (it is our 
understanding that those operators are meant for Euclidean, rather than
Lorentzian gravity). The only thing they share is that the square of 
the operators in $\cite{12,13}$, which is singular, and \cite{0},(3.10)
possess classical limits which are proportional to each other. It is 
therefore to be expected that the solutions that have been
found already in the literature for the formal square of those 
operators in \cite{12,13} (see, for instance, \cite{26,27})
are far from being annihilated by our operator. 
What is appealing 
about the operators constructed here is that they present quantizations of
\cite{0}, (2.1), the original Wheeler-Dewitt constraint, rather than the 
square root of a rescaled version thereof.\\
$\bullet$ Interestingly, although the classical theory only makes sense 
for non-degenerate metrics, the quantum theory does not blow up on states
which represent degenerate metrics since the volume operator only occurs
in a positive power. 
While this has been shown to be possible also in the 
Ashtekar framework \cite{3} (that is, after rescaling by 
$\sqrt{\det(q)}$) we 
see this effect already in the original framework without rescaling. \\
$\bullet$ There is a lot of freedom involved in the regularization 
step reflecting the fact that the quantum theory of a given classical field
theory is not unique. An important but unresolved question is how to
select the correct (physically relevant) regularization procedure.
A possible avenue to resolve this question is to apply the framework 
to exactly soluble models and to compare the results.\\
Another interesting question is how much freedom there actually
remains in the regularization step once we imposed our requirements as 
stated in section 3.1.2 of \cite{0}.\\
$\bullet$ The final expression of the Wheeler-DeWitt constraint 
is surprisingly simple : on each cylindrical function it is a low order
polynomial in the volume operator and holonomy operators and therefore
one can find exact solutions to the Quantum Einstein Equations, perhaps 
even easier than it is possible to find classical solutions. 
Remarkably, the spectrum of the Hamiltonian constraint operator at a given 
vertex is largely determined by the spectrum of the volume operator so that it
becomes important to gain control over it \cite{ThieVol}. \\
$\bullet$ Our simple trick, which essentially consists in replacing 
$e_a^i$ by
$\{A_a^i,V\}$ and so renders the seemingly ill-defined, non-polynomial,
non-analytic (in $E^a_i$) operator $\hat{e}_a^i$ into a perfectly 
well-defined quantity can also be applied to making sense out of operators 
which so far were completely out of reach as they involve $q_{ab}$
and thus cannot be written as square roots of polynomials in $E^a_i$.
This class of operators includes, but does not exhaust, operators that 
measure the length of a curve \cite{14}, the quantum generators of the
asymptotic Poincar\'e group \cite{14a} and Hamiltonian operators describing 
the matter sector, as for instance Yang-Mills theory \cite{15}.\\
$\bullet$ Concluding, we have shown, that there exists a mathematically 
rigorous and consistent way
to non-perturbatively quantize the Lorentzian Wheeler-DeWitt constraint for 
full four-dimensional vacuum gravity in the continuum. The stage is set to
solve the theory, that is, to find explicitly the physical states, 
observables and to compute their spectra. As outlined above, 
modulo 
computing the precise coefficients of the expansion of a solution in 
terms of diffeomorphism invariant spin-network states (we also have given a 
method of computation), at least for the 
non-symmetric operator we already computed the physical Hilbert space.
We are now in the position to settle non-perturbatively and rigorously
questions that arise, for instance, in black hole physics.
\end{itemize}

{\large Acknowledgments}\\
\\
This research project was 
supported in part by DOE-Grant DE-FG02-94ER25228 to Harvard University.


\begin{thebibliography}{99}

\parskip -5pt

\bibitem{0} T. Thiemann, ``Quantum Spin Dynamics (QSD)", 
Harvard Preprint HUTMP-96/B-351, The previous paper in this volume
or in the gr-qc archive.

\bibitem{3} A.\ Ashtekar, Phys.\ Rev. Lett.\ {\bf 57} 2244 (1986),
            Phys.\ Rev.\ {\bf D36}, 1587 (1987).

\bibitem{4} T. Thiemann, Class. Quantum Grav. {\bf 13} (1996) 1383-1403

\bibitem{5} A. Ashtekar, ``A generalized Wick transform for gravity", 
preprint CGPG-95/12-1, gr-qc/9511083

\bibitem{7} A. Ashtekar and C.J. Isham,
Class. \& Quan. Grav. {\bf 9}, 1433 (1992).

\bibitem{8} A. Ashtekar and J. Lewandowski, ``Representation
theory of analytic holonomy $C^\star$ algebras'', in {\it Knots and
quantum gravity}, J. Baez (ed), (Oxford University Press, Oxford 1994)

\bibitem{9} J. Baez, Lett. Math. Phys. {\bf 31}, 213 (1994);
``Diffeomorphism invariant generalized measures on the space of
connections modulo gauge transformations", hep-th/9305045,
in the Proceedings of the conference on quantum topology, D. Yetter
(ed) (World Scientific, Singapore, 1994).

\bibitem{11} A. Ashtekar, J. Lewandowski, Journ. Geo. Physics {\bf 17} 
(1995) 191, J. Math. Phys. {\bf 36}, 2170 (1995) 

\bibitem{18} A. Ashtekar, J. Lewandowski, D. Marolf, J. Mour\~ao, T.
Thiemann, ``Quantization for diffeomorphism invariant theories 
of connections with local degrees of freedom", Journ. Math. Phys.
{\bf 36} (1995) 519-551

\bibitem{12} C. Rovelli, L. Smolin, Phys. Rev. Lett. {\bf 72} (1994) 446

\bibitem{13} A. Ashtekar, J. Lewandowski, ``Regularization of the 
Hamiltonian constraint", in preparation

\bibitem{14} T. Thiemann, ``The length operator in canonical quantum 
gravity", Harvard Preprint HUTMP-96/B-394

\bibitem{14a} T. Thiemann, ``A Hamiltonian operator for canonical quantum 
gravity", Harvard University Preprint

\bibitem{15} T. Thiemann, ``A regularization of canonical Yang-Mills quantum
theory", Harvard University Preprint

\bibitem{16} A. Higuchi Class. Quant. Grav. {\bf 8},  1983 (1991),
Class. Quant. Grav. {\bf 8}, 2023 (1991)

\bibitem{17} D. Marolf, ``The spectral analysis inner product for
quantum gravity,'' preprint gr-qc/9409036, to appear in the
Proceedings of the VIIth Marcel-Grossman Conference, R. Ruffini and
M. Keiser (eds) (World Scientific, Singapore, 1995),
Class. Quant. Grav. (1995) \\
``Almost Ideal Clocks in Quantum Cosmology: A Brief
Derivation of Time,'' preprint gr-qc/9412016.

\bibitem{BaSa} J. Baez, S. Sawin, ``Functional Integration on Spaces of 
Connections", q-alg/9507023

\bibitem{24} M. Reed, B. Simon, ``Functional Analysis",
Mod. Meth. Math. Phys. Vol II, Avcademic Press, New York, 1970

\bibitem{26} B. Br\"ugmann, J. Pullin, Nucl. Phys. B {\bf 363} 221

\bibitem{27} B. Br\"ugmann, J. Pullin, R. Gambini, Phys. Rev. Lett. {\bf 68} 
(1992) 431, Nucl. Phys. B {\bf 385} (1992) 1199

\bibitem{27a} T. Thiemann, ``Spectral Analysis of the Wheeler-DeWitt
constraint operator", Harvard University preprint

\bibitem{28} R. Loll, ``A real alternative to to quantum gravity in loop
space", Preprint DFF 244/02/96, gr-qc/9602041

\bibitem{29} T. Thiemann, ``The ADM Hamiltonian operator for canonical 
quantum gravity", Harvard University Preprint

\bibitem{ThieVol} T. Thiemann, ``Complete formula for the matrix elements 
of the volume 
operator in canonical quantum gravity", Harvard Preprint
HUTMP-96/B-393

\end{thebibliography}
\end{document}